\documentclass[a4paper,12pt]{article}

\usepackage{amsmath,amssymb,amscd,mathtools,array,empheq}
\usepackage[utf8]{inputenc}
\usepackage{lmodern,newtxtext,newtxmath}
\usepackage[a4paper,tmargin=3truecm,bmargin=3truecm,hmargin=1.6truecm]{geometry}
\usepackage[colorlinks=true, urlcolor=blue, citecolor=blue, linkcolor=blue,pdfusetitle]{hyperref}
\usepackage{graphicx}
\usepackage{color}
\usepackage[dvipsnames]{xcolor}
\usepackage[english]{babel}
\usepackage{bm}
\usepackage{setspace}
\usepackage{cite}
\usepackage{etoolbox}
\apptocmd{\thebibliography}{\setlength{\itemsep}{0pt}}{}{}
\usepackage[shortlabels]{enumitem}
\numberwithin{equation}{section}

\usepackage{tikz}
\usetikzlibrary{shapes,arrows,arrows.meta,matrix,decorations.markings,calc,babel,quotes,angles}
\makeatletter \g@addto@macro\@floatboxreset\centering \makeatother

\DeclareMathOperator{\Ad}{Ad}
\DeclareMathOperator{\Coad}{Coad}
\DeclareMathOperator{\const}{const}

\DeclareMathOperator{\Diff}{Diff}

\newcommand{\R}{\mathbb{R}}
\newcommand{\C}{\mathbb{C}}
\newcommand{\Z}{\mathbb{Z}}

\newcommand{\la}{{\langle}}
\newcommand{\ra}{{\rangle}}

\newcommand{\Carr}{\mathrm{Carr}}
\newcommand{\eCarr}{\widetilde{\mathrm{Carr}}}

\newcommand{\vol}{\mathrm{vol}}
\newcommand{\SL}{\mathrm{SL}}

\newcommand{\GL}{\mathrm{GL}}
\newcommand{\SO}{\mathrm{SO}}
\newcommand{\SE}{\mathrm{SE}}
\newcommand{\ecarr}{\widetilde{\mathfrak{carr}}}
\newcommand{\carr}{\mathfrak{carr}}
\newcommand{\so}{\mathfrak{so}}
\newcommand{\gl}{\mathfrak{gl}}

\newcommand{\ie}{\textit{i.e.} }
\newcommand{\eg}{\textit{e.g.} }
\newcommand{\half}{\frac{1}{2}}

\newcommand{\rg}{\mathrm{g}}

\newcommand{\cF}{\mathcal{F}}

\newcommand{\cH}{\mathcal{H}}
\newcommand{\cI}{\mathcal{I}}

\newcommand{\cN}{\mathcal{N}}
\newcommand{\cO}{\mathcal{O}}

\newcommand{\cT}{\mathcal{T}}

\newcommand{\bomega}{\boldsymbol{\omega}}

\newcommand{\bbeta}{\boldsymbol{\beta}}
\newcommand{\bgamma}{\boldsymbol{\gamma}}

\newcommand{\bnabla}{\boldsymbol{\nabla}}
\newcommand{\ba}{\boldsymbol{a}}
\newcommand{\bb}{\boldsymbol{b}}
\newcommand{\bc}{\boldsymbol{c}}
\newcommand{\bg}{\boldsymbol{g}}
\newcommand{\bl}{\boldsymbol{l}}
\newcommand{\bp}{{\boldsymbol{p}}}
\newcommand{\bu}{\boldsymbol{u}}
\newcommand{\bv}{{\boldsymbol{v}}}
\newcommand{\bx}{{\boldsymbol{x}}}

\newcommand{\bE}{\boldsymbol{E}}
\newcommand{\bB}{\boldsymbol{B}}
\newcommand{\bzero}{\boldsymbol{0}}
\newcommand{\spin}{\mathtt{s}}

\newcommand{\womega}{\widetilde{\omega}}

\newcommand{\tm}{\widetilde{m}}

\usepackage{scalerel}
\let\savewidetilde\widetilde
\def\widetilde#1{%
 \ThisStyle{\savewidetilde{\phantom{\SavedStyle#1}}%
  \setbox0=\hbox{$\SavedStyle#1$}\kern-\wd0#1}}

\title{Planar Carrollean dynamics, and the Carroll quantum equation}
 
\author{L. Marsot\footnote{mailto: loic.marsot@univ-amu.fr}\\[1.em]
{\normalsize Centre de Physique Théorique}\\
{\normalsize Aix Marseille Univ, Université de Toulon, CNRS, CPT, Marseille, France.}
} 

\date{{\footnotesize (\today)}}

\begin{document}

\maketitle

\bigskip

\begin{abstract}
We expand on the known result that the Carroll algebra in $2+1$ dimensions admits two non-trivial central extensions by computing the associated Lie group, which we call extended Carroll group. The symplectic geometry associated to this group is then computed to describe the motion of planar Carroll elementary particles, in the free case, when coupled to an electromagnetic field, and to a gravitational field. We compare to the motions of Carroll particles in $3+1$ dimensions in the same conditions, and also give the dynamics of Carroll particles with spin. In an electromagnetic background, the planar Carroll dynamics differ from the known Carroll ones due to 2 new Casimir invariants, and turn out to be non-trivial. The coupling to a gravitational field leaves the dynamics trivial, however. Finally, we obtain the quantum equation obeyed by Carroll wave functions \textit{via} geometric quantization.
\end{abstract}

\section{Introduction}

In the late 1960s, the possible kinematical groups were classified \cite{BacryLL68,LevyLeblond76}, assuming isotropy and homogeneity of spacetime, and a ``weak'' causality condition. Among them, alongside notably the Poincaré group and the Galilei group, was the Carroll group. This group was discovered a few years earlier as an ``ultrarelativistic'' contraction of the Poincaré group \cite{LevyLeblond65} (often said to be the limit $c \rightarrow 0$), in contrast to the Galilei group which is a ``non relativistic'' contraction ($c \rightarrow \infty$). Thus, while the Galilei contraction of the Poincaré group ``opens up'' its light-cone structure, the light cone structure of the Carroll group collapses into a line along the time axis.

Both the Galilei group and the Carroll group feature rotations, space time translations and boosts. However, the boosts act on space for the former, and on time for the latter. Another characteristic, or rather lack of, of the Carroll group is that it does not admit non-trivial central extensions in dimensions $3+1$ and higher. This is unlike the Galilei group, which always admits a non-trivial central extension \cite{LevyLeblond71}. This is important, because while the Galilei group features intrinsically the conservation of energy, through its time translation symmetry, it gains the conservation of the mass of elementary particles through its central extension. For the Carroll group, however, it has intrinsically mass conservation, but there is no central extension to conserve the energy.

The Carroll group has given birth to Carroll \emph{structures} \cite{Henneaux79,DuvalGH91,Dautcourt98,DuvalGH14,DuvalGHZ14} which are, with some abuse of language, the dual construction of Newton-Cartan structures. Recall that Newton-Cartan structures depict the geometry of non-relativistic (or rather, Galilei) spacetime \cite{Cartan23,Trautman63,Havas64,Trautman67,Kunzle72}.

\medskip

A Carroll structure is defined as a triple $(M, g, \xi)$ consisting of,
\vspace{-0.5em}
\begin{enumerate}[i)]
\setlength\itemsep{-0.3em}
\item a manifold $M$ of dimension $d+1$ ;
\item a degenerate, twice symmetric, covariant tensor $g$, such that $\dim \ker g = 1$ ;
\item a nowhere vanishing vector field $\xi \in \ker g$ ;
\item together with the compatibility condition that the Lie derivative of $g$ along $\xi$ vanishes\footnote{It is possible to relax this condition, which is somewhat analogous to the closure of the clock 1-form on Newton-Cartan structures, here, as our paper does not depend on this, thus allowing for a slight generalization.}, $L_\xi g = 0$.
\end{enumerate}
It is possible to extend such structures to include a (non unique) connection, so that we have the quadruple $(M, g, \xi, \nabla)$, where the connection is compatible with both the ``metric'' $g$ and the vector field $\xi$, \ie $\nabla g = 0$ and $\nabla \xi = 0$. 

The first obvious example of a Carroll structure is the structure obtained as the Carrollean limit of a Minkowski spacetime\footnote{This is achieved by defining the time coordinate as $x^4 := s/C$, in contrast to $x^4 := c t$ for the Galilean limit, and letting $C \rightarrow \infty$. The metric becomes degenerate, and $\xi = \partial_s$ is in its kernel. The Lie derivative condition is then trivial in the flat case.}. There are more physically interesting examples, however. It has been shown in \cite{Morand18,CiambelliLMP19} that embedded null hypersurfaces in a Lorentzian spacetime are Carroll surfaces. An example of these is the horizon of black holes \cite{DonnayM19}. Another example of Carroll structure is null infinity. Indeed, it is immediate that the definition of null infinity $\cI^\pm$, given in \eg \cite{Ashtekar14} (a few lines above the eq. (2.4)), satisfies all the definition items of a Carroll structure above. This example is of particular relevance, as null infinity has been shown to have BMS symmetry \cite{BondiBM62,Sachs62}, and one can recover the BMS group as the group of conformal isomorphisms of the Carroll structure $(S^2 \times \R, g, \xi) \cong \cI^\pm$, \cite{DuvalGH14}.

A fourth example, which is another example of embedded null hypersurface, but which will be relevant in this paper, are those Carroll structures obtained as a $t = \const$ slice of a Bargmann structure \cite{Bargmann54,DuvalBKP85,Eisenhart28}, which is a principal $\R$ (or $S^1$)-bundle over a Newton-Cartan structure, with the aim of describing Galilean physics in a covariant way. See figure \ref{f:barg}.

\newcounter{w}
\newcounter{h}
\newcounter{p}
\setcounter{w}{9}
\setcounter{p}{1}
\setcounter{h}{7}

\tikzstyle{hidden} = [dashed,line width=1.1pt]
\tikzstyle{lesser} = [line width=1.2pt]
\tikzstyle{normal} = [line width=0.8pt]
\tikzstyle{normalh} = [dashed,line width=0.8pt]
\tikzstyle{arrow} = [line width=0.9pt, draw, -latex']
\tikzstyle{cone} = [line width=0.7pt]
\tikzstyle{labels} = [->]
\tikzstyle{carr} = [black!50!blue]
\tikzstyle{line} = [draw, -latex']
\tikzstyle{nc} = [black!50!red]

\tikzset{middlearrow/.style={
        decoration={markings,
            mark= at position #1 with {\arrow{{>}[scale=1.5]}} ,
        },
        postaction={decorate}
    }
}

\begin{figure}[ht]
\begin{tikzpicture}[line width=1.4pt,scale=0.85, every node/.style={transform shape}]
  \draw [lesser] (0,0) -- (0.45 * \value{w},\value{p}) -- (\value{w},0);
  \draw (\value{w},0) -- (0.55 * \value{w},-\value{p}) -- (0,0);

  \draw [line width=1.4pt] (0,0) -- (0,-\value{h});
  \draw (0.55 * \value{w}, -\value{p}) -- (0.55 * \value{w}, -\value{p} - \value{h});
  \draw (\value{w}, 0) -- (\value{w}, -\value{h});
  \draw [hidden] (0.45 * \value{w},\value{p}) -- (0.45 * \value{w},\value{p} - \value{h});
  \draw [normal] (0.9 * \value{w}, - \value{p}) circle (0.6cm) node [scale=1.5]{$M$};

  \draw [hidden] (0, - \value{h}) -- (0.45 * \value{w},\value{p} - \value{h}) -- (\value{w}, - \value{h});
  \draw (\value{w}, - \value{h}) -- (0.55 * \value{w},-\value{p} - \value{h}) -- (0, - \value{h});

  \draw [lesser,nc] (0, - 1.4 * \value{h}) -- (0.45 * \value{w},\value{p} - 1.4 * \value{h}) -- (\value{w}, - 1.4 * \value{h});
  \draw [nc] (\value{w}, - 1.4 * \value{h}) coordinate (nc1) -- node[pos=1,above,scale=1.5]{$\cN$} (0.55 * \value{w},-\value{p} - 1.4 * \value{h}) coordinate (nc2) -- (0, -1.4 * \value{h}) coordinate (nc3);
  \draw [normal] pic["",draw=black,-,angle eccentricity=1.2,angle radius=0.85cm] {angle=nc1--nc2--nc3};

  \draw [normal,carr] (0.725 * \value{w}, 0.5*\value{p}) -- (0.275 * \value{w}, -0.5*\value{p}) coordinate (c1);
  \draw [normal,carr] (0.275 * \value{w}, -0.5*\value{p}) -- node[pos=0.92,right,scale=1.5]{$\widetilde{\Sigma}_t$} (0.275 * \value{w}, -1*\value{h} -0.5*\value{p});
  \draw [normalh,carr] (0.725 * \value{w}, 0.5*\value{p}) -- (0.725 * \value{w}, 0.5*\value{p} - \value{h}) coordinate (c3) -- (0.275 * \value{w}, -\value{h}-0.5*\value{p}) coordinate (c2);
  \draw [normal] pic["",draw=black,-,angle radius=1.25cm] {angle=c3--c2--c1};

  \draw [normal, middlearrow={0.82},black!60!green] (0.725 * \value{w}, 0.5 * \value{p} - 1.4 * \value{h}) -- node [left,pos=0.32,scale=1.1] {$(x, t)\qquad$} node[scale=3.5,pos=1]{.} (0.05 * \value{w}, -1.4 * \value{h} - 1 * \value{p});
  
  \draw [line] (-0.225 * \value{w}, -1.4 * \value{h} - 0.5* \value{p}) -- node [pos=1.18,below,scale=1.1] {$T \cong \R$ (time axis)} node[below, pos=0.58,scale=1.1]{$t = \const \qquad$} (0.325 * \value{w}, -1.5 * \value{p} - 1.4 * \value{h});
  
  \draw [arrow] (0.80 * \value{w}, -0.65 * \value{h}) -- node [near end, left, scale=1.1] {$\xi$} (0.80 * \value{w}, -0.53 * \value{h});
  \draw [arrow] (0.40 * \value{w}, -0.4 * \value{h}) -- node[pos=0,scale=3]{.} node [left,pos=0,scale=1.1]{$(x,t,s)$} node [near end, right,scale=1.1] {$\xi$} (0.40 * \value{w}, -0.28 * \value{h});

  \draw [dashed,cone] (0.80 * \value{w}, -0.65 * \value{h}) -- (0.80 * \value{w}, -0.77 * \value{h});
  \draw [dashed,cone] (0.80 * \value{w}, -0.65 * \value{h}) -- (0.745 * \value{w}, -0.74 * \value{h});
  \draw [cone] (0.80 * \value{w}, -0.65 * \value{h}) -- (0.855 * \value{w}, -0.56 * \value{h});
  \draw [cone,rotate around={160:(0.827 * \value{w}, -0.55 * \value{h})}] (0.827 * \value{w}, -0.55 * \value{h}) ellipse (0.029 * \value{w} and 0.02 * \value{h});
  \draw [cone,dashed,rotate around={160:(0.773*\value{w}, -0.75 * \value{h})}] (0.773*\value{w}, -0.75 * \value{h}) ellipse (0.0295 * \value{w} and 0.02 * \value{h});
  
  \node[draw, align=center,scale=1.1] at (1.17*\value{w}, \value{p}) (barg) {Bargmann\\space-time\\$(M, \widetilde{\rg}, \widetilde{\xi})$};
  \draw [labels] (barg) -- (0.83 * \value{w}, -0.1 * \value{p});
  
  \node[draw=black!50!blue, align=center,scale=1.1] at (-0.15 * \value{w}, \value{p}) (carr) {Carroll\\space-time\\$(\widetilde{\Sigma}_t, g, \xi)$};
  \node[draw=none] at (0.4 * \value{w}, -0.5*\value{p}) (carr2) {};
  \draw [labels] (carr) to [out=0,in=100] (carr2);
  
  \node[draw=black!60!green, align=center,scale=1.1] at (1*\value{w}, -\value{h} - 1.5*\value{p}) (euclide) {Euclidean\\space\\$(\Sigma_t, h)$};
  \node[draw=none] at (0.55*\value{w}, -1*\value{h}-2.83*\value{p}) (euclide2) {};
  \draw [labels] (euclide) to [out=180,in=45] (euclide2);
  
  \node[draw=black!50!red, align=center,scale=1.1] at (1.1 * \value{w}, -\value{h} - 4.1 * \value{p}) (nc) {Newton-Cartan\\space-time\\$(\cN, h, \theta, \nabla^\cN)$};
  \node[draw=none] at (0.8 * \value{w}, -\value{h} - 2.95 * \value{p}) (nc2) {};
  \draw [labels] (nc) to [out=140,in=25] (nc2);
  
  \draw [normalh, middlearrow=0.4] (0.40 * \value{w}, -0.4 * \value{h}) -- (0.40 * \value{w}, -\value{h} - 0.75 * \value{p});
  \draw [normal, middlearrow=0.36] (0.40 * \value{w}, -\value{h} - 0.75 * \value{p}) -- node [left,pos=0.32,scale=1.5]{$\pi$} node [pos=1,scale=3]{.} (0.40 * \value{w}, -1.12 * \value{h} - 2.22 * \value{p});
\end{tikzpicture}
\caption{Visualization of a 1+2 dimensional Bargmann structure, and its link to Newton-Cartan and Carroll structures.}
\label{f:barg}
\end{figure}

Moreover, the Carroll group itself has seen some use in the recent literature. For instance in  \cite{DuvalGHZ17}, the well-known isometry group of gravitational waves has been identified to be the subgroup of the Carroll group without rotations. Also, in \cite{BergshoeffGL14} the authors considered the dynamics of a system of Carroll particles, as well as gauged particles to obtain their behavior in a gravitational field. See also \cite{Trzesniewski18}. Let us finally mention that the Carroll group was applied in holography and string theory \cite{BagchiBKM16,CardonaGP16}.

As we have seen a few paragraphs above, Carroll structures of dimension $2+1$ are of particular relevance (more so than those of dimension $3+1$, even, given that there are physical examples of planar Carroll structures) and, quite interestingly, the Carroll group in $2+1$ dimensions has been found to admit a non-trivial central extension of dimension 2 \cite{NgendakumanaNT14,Ngendakumana14,AzcarragaHPS98}, much like the Galilei group, see \eg \cite{LevyLeblond71,BallesterosGD92}. This central extension fact has mostly been missed, or forgotten about, in the recent literature about the dynamics of Carroll particles. 

Describing the dynamics of Carroll elementary particles means to write down equations of motion that the particles follow, equations which in turn can be described by a Carroll-homogeneous symplectic manifold. Now, by the (converse of the) Kirillov-Kostant-Souriau theorem, the associated symplectic manifold is locally symplectomorphic to a coadjoint orbit of the Carroll group, or a non-trivial central extension of this group. The equations of motion for Carroll particles given in, \eg \cite{DuvalGHZ14}, are valid in $3+1$ dimensions or higher because in this case the group does not admit central extensions (as already stated by the authors of \cite{DuvalGHZ14}), but may not be valid in $2+1$ dimensions, since there are central extensions to be taken into account. As a matter of fact, in the latter case, one should consider the equations of motion spanned by the non-trivial \emph{central extension} of the Carroll group. In particular, the complete description of a planar Carroll elementary particle involves two additional Casimir invariants. Recall that considering the central extension is indeed important. For instance, if one forgets about the central extension of the Galilei group (in any dimension), then the resulting equations of motion may only describe massless particles, since the mass arises as the Casimir invariant obtained from the central extension of the Galilei group.

The paper is organized as follows. We will recall the definition of the Carroll group and its properties in the section~\ref{s:carr_group}, as well as compute the group of the double central extension of the Carroll group in $2+1$ dimensions from the algebra computed in \cite{NgendakumanaNT14,Ngendakumana14,AzcarragaHPS98}. The planar version of the group is especially important in Carrollean dynamics owing to the above mentioned fact that a null hypersurface embedded in a $3+1$ Lorentzian manifold is a Carroll structure \cite{Morand18,CiambelliLMP19}.

The aim of the section~\ref{s:motions} is to describe, using symplectic geometry, the dynamics of Carroll elementary particles in $3+1$ dimensions and $2+1$ dimensions, accounting for the non-trivial central extensions in the latter case. We will describe the free case, the coupling to electromagnetism, and the coupling to a gravitational field. We also compute the motions of Carroll particles with spin. 

Then, thanks to the symplectic models computed in the previous section, we find the quantum equation describing free Carroll wavefunctions, with several methods, including geometric quantization, in section~\ref{s:quantum}.

\section{Carroll-related groups}
\label{s:carr_group}

\subsection{The Carroll group}

As recalled in the introduction, the Carroll group and algebra can be obtained from a In\"on\"u-Wigner contraction \cite{InonuW53} of the Poincaré group, as shown by Lévy-Leblond \cite{LevyLeblond65}. This contraction corresponds to taking the limit $c \rightarrow 0$, in opposition to the limit $c \rightarrow \infty$ which leads to the Galilei group \cite{InonuW53}. Note that in practice, one does not directly take the limit $c \rightarrow 0$. It is instead more convenient to define a velocity $C$ such that the time-like coordinate on a Lorentzian manifold is defined as $x^4 := s/C$ (as opposed to $x^4 := c t$), and then take the limit $C \rightarrow \infty$. This has some important implications, however, since $C$ is defined to have the dimensions of a velocity, the ``time'' variable $s$ now has the dimension of an action per mass, \ie $L^2 T^{-1}$. 

The Carroll group, denoted $\Carr(d+1)$ is isomorphic to the subgroup of $\GL(d+2, \R)$ of elements $a_V \in \GL(d+2, \R)$\footnote{The notation $a_V$ is for the representation of the group element $a$ as a linear map acting on the representation space $V$.},
\begin{equation}
\label{gp_mat_carroll}
a_V = \left(\begin{array}{ccc}
A & 0 & \bc \\
- \overline{\bb} A & 1 & f \\
0 & 0 & 1
\end{array}\right)
\end{equation}
with $A \in \SO(d)$ a rotation, $\bb \in \R^d$ a boost, $\bc \in \R^d$ a space translation, and $f \in \R$ a ``time'' translation, and where the bar denotes the transposition in $\R^d$ with respect to spatial part of the metric. This group acts projectively on $\R^{d+1}$, or linearly on the representation space $V = \R^{d+1} \times \{1\}$,
\begin{equation}
\left(\begin{matrix}
\bx \\
s \\
1
\end{matrix}\right)
\mapsto
\left(\begin{matrix}
A \bx + \bc \\
s - \la \bb,  A \bx \ra + f \\
1
\end{matrix}\right).
\end{equation}

The main difference between the Galilei group and this group is that instead of the boosts acting on the spatial coordinates, they act on the time coordinate.

The Carroll group gives its name to Carroll structures, whose definition can be found in the introduction. Indeed, the isometry group of a flat Carroll structure $(\R^{d,1}, \delta, \xi, \nabla)$, \ie the group such that $\Phi^*\delta = \delta, \Phi^* \xi = \xi, \Phi^* \nabla = \nabla$, for $\Phi \in \Carr(d+1)$, and where $\delta$ is the flat spatial metric, is isomorphic to the Carroll group. Note that preserving the connection is required to reduce the isometries of the flat structure $(\R^{d,1}, \delta, \xi)$, which are infinite-dimensional due to the degeneracy of $g$, to the Carroll group. Recall that the same phenomenon happens with flat Newton-Cartan structures: the (contravariant) metric is degenerate, and thus the group of isometries is infinite dimensional. It is only when asking for the (non unique, again) connection to be preserved that one ends up with the Galilei group.

The generators of the Lie algebra $\carr(d+1)$ are $(J_i), (P_i), (K_i), (M)$ of, respectively, rotations, spatial translations, boosts, and time translations, with non trivial commutators ($d = 3$ is implied here),
\begin{equation}
\label{alg_carr}
\begin{array}{cccc}
[J_i, J_j] = \epsilon_{ijk} J_k, & [J_i, P_j] = \epsilon_{ijk} P_k, & [J_i, K_j] = \epsilon_{ijk} K_k, & [K_i, P_j] = M \delta_{ij}
\end{array}
\end{equation}

This Lie algebra is isomorphic to the space of vector fields $X_V \in T_y M$, $y = (\bx, s, 1) \in V$,
\begin{equation}
\label{alg_vect_carroll}
X_V = \left(\overline{j(\bomega) \bx} + \overline{\bgamma}\right) \partial_\bx  + \left(- \la\bbeta, \bx\ra + \varphi\right) \partial_s,
\end{equation}
where $j$ is the linear map $j: \R^{d(d-1)/2} \rightarrow \so(d)$, $\bgamma \in \R^d$ is a space translation, $\boldsymbol{\beta} \in \R^d$ is a boost, and $\varphi \in \R$ is a time translation, together with the commutator of vector fields.

This algebra can also be represented as a subalgebra of $\gl(d+2, \R)$, with elements $Z_V \in \gl(d+2)$,
\begin{equation}
\label{alg_mat_carroll}
Z_V = \left(\begin{array}{ccc}
j(\bomega) & 0 & \gamma \\
- \overline{\bbeta} & 0 & \varphi \\
0 & 0 & 0
\end{array}\right).
\end{equation}

\subsection{Carroll for \texorpdfstring{$S^2 \times \R$}{S²xR}}

As mentioned in the introduction, an interesting class of Carroll structures, important for their physical relevance, are those where the base manifold is $S^2 \times \R$, namely the horizon of black holes and null infinity. We will hence denote the structure as the triple $(S^2\times \R, g, \xi)$ where the degenerate metric is locally $g = g_\Sigma + 0 \cdot ds$, where $g_\Sigma$ is the metric of the 2-sphere, so that $g(\xi) = 0$, with $\xi = \partial_s$.\footnote{For instance, it was explicitly shown in \cite{Marsot16} that the horizon of a Kerr-Newman black hole is such a Carroll structure. Indeed, let us take the Kerr-Newman metric, with $\Delta = r^2 - 2 M r + a^2 + Q^2$, and $\Sigma = r^2 + a^2 \cos^2 \theta$,
\begin{equation*}
g = - \frac{\Delta}{\Sigma} \left(dt - a \sin^2 \theta \, d\varphi\right)^2 + \frac{\sin^2 \theta}{\Sigma} \left(a \, dt - (r^2+a^2) d\varphi\right)^2 + \Sigma d\theta^2 + \frac{\Sigma}{\Delta} dr^2,
\end{equation*}
and compute the induced metric on the horizon, defined at $\Delta = 0$ with $r = \const$. One obtains the degenerate metric,
\begin{equation*}
\widetilde{g} = \frac{\sin^2 \theta}{\Sigma} \left(a \, dt - (r^2+a^2) d\varphi\right)^2 + \Sigma d\theta^2,
\end{equation*}
whose kernel is spanned by the vector field $\xi$,
\begin{equation*}
\xi = \partial_t + \frac{a}{r^2+a^2} \partial_\varphi.
\end{equation*}

To make the Carroll structure clearer, one can change coordinates $(\theta, \varphi, t) \mapsto (\theta, \widetilde{\varphi} = \varphi - \frac{a}{r^2+a^2} s, s = t)$ such that we have,
\begin{equation}
\label{BHH}
\widetilde{g} = \frac{(r^2+a^2) \sin^2 \theta}{\Sigma} d\widetilde{\varphi}^2 + \Sigma d\theta^2 \quad \& \quad \xi = \partial_s.
\end{equation}

It is then immediate that \eqref{BHH} fulfills the definition of a Carroll structure.
}

The group of automorphisms of $(S^2 \times \R, g, \xi)$, \ie the subgroup of diffeomorphisms $\Phi \in \Diff(S^2\times\R)$ such that $\Phi^* g = g$ and $\Phi^* \xi = \xi$ is infinite dimensional. It is readily seen to be $\SO(3) \ltimes \cT$, where $\cT = C^\infty(S^2, \R)$ are often called \emph{super translations} \cite{Sachs62}. This is closely related to the BMS group. Indeed, it has been shown in \cite{DuvalGH14} that conformal transformations of the Carroll structure $(S^2 \times \R, g, \xi)$ such that the ``universal structure'' $g \otimes \xi \otimes \xi$ is preserved, turn out to form the BMS group \cite{BondiBM62,Sachs62}, $\mathrm{BMS}(4) \cong \SL(2, \C) \ltimes \cT$. 

If one wants to reduce this group of isometries to a group of finite dimension, it is customary to ask for the preservation of a connection defined on $S^2\times \R$ together with the metric and vector field. The isometries of the Carroll structure $(S^2 \times \R, g, \xi, \nabla)$ are then reduced to $\SO(3) \times \R$. 

\subsection{A double central extension for Carroll in 2+1 dimensions}
\label{ss:ecarr}

While the cohomology of the Galilei and Carroll groups are distinct in $3+1$ dimensions and higher\footnote{The Galilei group always admits a non-trivial central extension for $d \ge 3$: the Bargmann group, while the Carroll group does not have non-trivial central extensions for $d \ge 3$.}, they share similar features in $2+1$ dimensions. It is well known that the Galilei group admits 2 non-trivial central extensions in dimension $2+1$\cite{LevyLeblond71,BallesterosGD92}, and the same turns out to be true for the Carroll group\footnote{A third \emph{kinematical} group which admits non-trivial extensions in $2+1$ dimensions is the Newton-Hooke group (``Newton group'' in \cite{BacryLL68}).}.

The doubly extended Carroll algebra is spanned by the generators $J_3$, $(P_i)$, $(K_i)$, and $M$ of the standard Carroll algebra \eqref{alg_carr} in $2+1$ dimensions, and also by two central parameters $(A_i)$, with $i = 1, 2$. Their non trivial commutators were computed in \cite{NgendakumanaNT14,Ngendakumana14,AzcarragaHPS98},
\begin{equation}
\label{alg_carr_ext}
\begin{array}{ccc}
[J_3, P_i] = \epsilon_{ij} P_j, & [P_i, P_j] = \epsilon_{ij} A_1, & [K_i, P_j] = M \delta_{ij}, \\ \relax
[J_3, K_i] = \epsilon_{ij} K_j, & [K_i, K_j] = \epsilon_{ij} A_2, & 
\end{array}
\end{equation}
where $\epsilon_{ij}$ are the components of the fully skew-symmetric tensor such that $\epsilon_{12} = 1$. We denote this algebra by $\ecarr(2+1)$.

Knowing the matrix representation of the Carroll algebra $\carr(d+1)$, see \eqref{alg_carr}, in $\gl(d+2,\R)$, see \eqref{alg_mat_carroll}, we easily find a (non irreducible) representation for the extended Carroll algebra $\ecarr(2+1)$ in $\gl(6,\R)$,
\begin{equation}
\label{alg_mat_ecarr}
Z_V = \left(\begin{matrix}
j(\omega) & 0 & \bgamma & 0 & \epsilon \bbeta \\
- \overline{\bbeta} & 0 & \varphi & 0 & \alpha_2 \\
0 & 0 & 0 & 0 & 0 \\
\overline{\bgamma} \epsilon & 0 & \alpha_1 & 0 & - \varphi \\
0 & 0 & 0 & 0 & 0
\end{matrix}\right),
\end{equation}
where $j(\omega) \in \so(2)$, $\bbeta \in \R^2$, $\bgamma \in \R^2$, $\varphi \in \R$ have the same meaning as for the Carroll algebra \eqref{alg_mat_carroll}, $\alpha_1$ and $\alpha_2$ are the coefficients related to the generators $A_1$ and $A_2$ respectively, and where $\epsilon$ without indices is understood to be the fully skew-symmetric matrix, $\epsilon = \left(\begin{matrix}
0 & 1 \\
-1 & 0 
\end{matrix}\right)$. Note that in 2 dimensions, we have $j(\omega) = \omega \epsilon$ \footnote{Also, the cross product of two vectors $\ba, \bb$ in 2 dimensions is the number $\ba \times \bb = \det(\ba, \bb) = \la\ba, \epsilon \bb\ra$.}, where $\omega \in \R$.

This representation can then be integrated to obtain a (non irreducible) representation of the extended group in $\GL(6,\R)$ thanks to the exponential map. Its elements are of the form,
\begin{equation}
\label{gp_mat_ecarroll}
\left(\begin{matrix}
A & 0 & \bc & 0 & \epsilon \bb \\
- \overline{B} A & 1 & f & 0 & a_2 \\
0 & 0 & 1 & 0 & 0 \\
- \overline{\epsilon \bc} A & 0 & a_1 & 1 & - (f + \la \bb, \bc\ra) \\
0 & 0 & 0 & 0 & 1
\end{matrix}\right).
\end{equation}

This representation makes the computation of the group law of $\eCarr(2+1)$ straightforward,
\begin{equation}
\label{gp_law_ecarroll}
\begin{array}{l}
(A, \bb, \bc, f, a_1, a_2) \cdot (A', \bb', \bc', f', a_1', a_2') = \\[0.6ex]
\qquad \qquad  \left(A A', A \bb' + \bb, A \bc' + \bc, f + f' - \la \bb, A \bc' \ra, a_1 + a_1' - \la \epsilon \bc, A \bc' \ra, a_2 + a_2' - \la \bb, A \epsilon \bb'\ra\right)
\end{array}
\end{equation}

\smallskip

Next, we want the coadjoint representation of the group on its algebra. To this end, define a moment $\mu$ in the dual of the Lie algebra, \ie $\mu = (l, \bg, \bp, m, q_1, q_2) \in \ecarr(2+1)^*$, together with the dual pairing, for $Z = (\omega, \bbeta, \bgamma, \varphi, \alpha_1, \alpha_2) \in \ecarr(2+1)$,
\begin{equation}
\label{pairing_ecarr}
\mu \cdot Z := l \omega - \la \bbeta, \bg \ra + \la \bgamma, \bp \ra + m \varphi + \alpha_1 q_1 + \alpha_2 q_2
\end{equation}

The coadjoint action of $a \in \eCarr(2+1)$ on the moment $\mu$ is defined through the usual formula $(\Coad(a) \mu) \cdot Z = \mu \cdot (\Ad(a^{-1}) Z)$. Given the group law and the pairing above, we find, for $a = (A, \bb, \bc, f, a_1, a_2) \in \eCarr(2+1)$, $\Coad(a) \mu = (l', \bg', \bp', m, q_1, q_2)$, with,
\begin{subequations}
\begin{empheq}[left=\empheqlbrace]{align}
l' & = l + \bb \times A \bg - \bc \times A \bp + m \bb \times \bc + q_1 \bc^2 - q_2 \bb^2 \\
\bg' & = A \bg + m \bc + 2 q_2 \epsilon \bb \label{coad_g}\\
\bp' & = A \bp + m \bb + 2 q_1 \epsilon \bc \label{coad_p}\\
m' & = m \\
q_1' & = q_1 \\
q_2' & = q_2
\end{empheq}
\end{subequations}

Quite interestingly, the two parameters $q_1$ and $q_2$ of the central extension mix, on the one hand, boosts $\bb$ with the moment $\bg$ \eqref{coad_g}, and on the other hand, translations $\bc$ with the momentum $\bp$ \eqref{coad_p}. There are four Casimir invariants under this coadjoint representation, which can be written as follows, if $m \neq 0$,
\begin{subequations}
\label{casimirs}
\begin{align}
C_1 & := m, \\
C_2 & := \left(1 + 4 \frac{q_1 q_2}{m^2}\right) l + \frac{\bg \times \bp}{m} + \frac{q_1}{m^2} \bg^2 - \frac{q_2}{m^2} \bp^2, \label{c2} \\
C_3 & := q_1, \\
C_4 & := q_2.
\end{align}
\end{subequations}

The Casimir invariants $C_1, C_3, C_4$ are, respectively, the mass, and the two charges associated to the central extensions. Since we are studying 2+1 dimensional systems, the second charge is related to the anyon spin \cite{Leinaas77,Wilczek82}. Note that $q_1$ has the physical dimensions of $MT^{-1}$ and $q_2$ of $MT$.

\section{Classical motions of Carroll particles in d+1 dimensions}
\label{s:motions}

\subsection{Dynamics of free Carroll particles}

\subsubsection{Dimension \texorpdfstring{$d \ge 3$}{d >= 3}}
\label{model_3d_free}

In the following, we are going to use the orbit method of symplectic geometry \cite{Kirillov62} to obtain the equations of motion. Recall that the space of (classical) solutions of a dynamical system (\eg the solutions of Newton's equations) is a symplectic manifold, see \eg \cite{Souriau70}. Now, by the converse of the Kirillov-Kostant-Souriau theorem, if this manifold is $G$-homogeneous (in the Newtonian case, $G$ would be the Galilei group) then it is locally symplectomorphic to a coadjoint orbit of $G$ or of a central extension of this group. Thus, from the study of the coadjoint orbits of the Carroll group, or its central extensions, we obtain the space of solutions to the associated dynamical system, and then the equations of motion.

Let us start with the simple case of a free Carroll particle where the group does not admit non-trivial central extensions, which happens when $d \ge 3$.  In this case, the trajectory of elementary particles can be identified with the co-adjoint action of the Carroll group on the moment describing the considered particle. 

Let us reinterpret the variables appearing in the Carroll group \eqref{gp_mat_carroll}, so that elements $a \in \Carr(d+1)$ may be represented as,
\begin{equation}
a_{\R^{d+1}\times \{1\}} = \left(\begin{array}{ccc}
A & 0 & \bx \\
- \overline{\bv} A & 1 & s\\
0 & 0 & 1
\end{array}\right),
\label{Carr}
\end{equation}
where we will interpret $(\bx,s)\in\R^{d+1}$ as a spacetime event and, naively for now, $\bv$ as linked to the momentum state of the particle\footnote{Just as in the Galilean framework, where one \emph{finds} that $d\bx/dt = \bv := \bp / m$, thus linking the \emph{velocity} to the \emph{momentum} of the particle.}. Note that the Carroll group may be viewed as the bundle of Carroll frames above spacetime $\R^{d+1}=\Carr(d+1)/\SE(d)$, the Euclidean group being parametrized by the couples $(A,\bv)$. Remember that the Carrollean ``time'' $s$ has the dimension of an action per mass.

We are going to consider two kinds of elementary particles: massive and spinless, and massive with spin. The dynamics of the first kind of elementary particles has already been studied in \cite{DuvalGHZ14}.

\paragraph{Spinless massive particles}

It was shown in \cite{DuvalGHZ14} that the dynamics of massive spinless particles is defined by the left-invariant $1$-form on $\Carr(d+1)$, 
\begin{equation}
\label{varpi_3d}
\varpi = m \la \bv, d\bx \ra + m ds
\end{equation}
where we interpret $\bp := m \bv$ as the momentum, and $m$ as the mass of the Carroll particle (one of the Casimir invariants of the group). Equation \eqref{varpi_3d} shows that this $1$-form, $\varpi$, actually descends to the \emph{evolution space}\footnote{Using the terminology of \cite{Souriau70}. \label{fn:terminology}} $V = \Carr(d+1) / \SO(d)\cong(T\R^d) \times \R$ above spacetime. The Carroll group \eqref{gp_mat_carroll} acts naturally on the evolution space, with $a \in \Carr(d+1)$ and $y = (\bx, \bv, s) \in V$ as,
\begin{equation}
\label{action_carr_v}
a_v (y) = (A \bx + \bc, A \bv + \bb, s - \overline{\bb} A \bx + f).
\end{equation}

The exterior derivative of the 1-form, 
\begin{equation}
\sigma := d\varpi = m d\overline{\bv} \wedge d\bx
\end{equation}
is presymplectic of rank $2d$. Indeed, its kernel provides the equations of motion. We have,
\begin{equation}
\label{kersigma_spinless}
\delta(\bx, \bv, s) \in \ker(\sigma) \; \Leftrightarrow  \; \delta\bx = 0, \delta \bv = 0, \delta{s} \in \R,
\end{equation}
and thus,
\begin{subequations}
\label{eom_free}
\begin{align}
\frac{d \bx}{ds} & = 0, \label{eom_free_dx}\\
\frac{d \bv}{ds} & = 0.
\end{align}
\end{subequations}

The quotient $U = V/\ker(\sigma) \cong T^*\R^d$ is called the \emph{space of motions}\footnote{\textit{Idem.}} of the model. This will be our symplectic manifold. It is, by construction, symplectomorphic to the $\Carr(d+1)$-coadjoint orbit of mass~$m\neq0$ and spin $\spin = 0$. It is clearly endowed with the symplectic $2$-form $\omega = d\overline{\bp} \wedge d\bx$, the image of $\sigma$ under the projection $V \to U$. As emphasized in \cite{DuvalGHZ14}, the dynamics of free massive spinless Carrollean particles are very poor as they do not move. Their worldlines are characterized by their absolute spatial location, $\bx \in \R^d$.

Notice that the equations of motion are \emph{not} $d\bx/ds = \bv$ with $\bv = 0$. While in the Galilean framework one finds the relation $d\bx/dt = \bv$, in the Carrollean case we find \eqref{eom_free_dx}, which means that we do not have a relation between the velocity and the momentum of the particle. Thus in the Carroll case, it is crucial to make the difference between the \emph{velocity} $d\bx/ds$ and the momentum state of the particle, which is linked to $\bv$.

\paragraph{Particles with spin}

Knowing the dynamics of massive spinless particles from the previous section, working out those of massive particles with spin is straightforward. Indeed, similarly to the case of Galilean particles with spin \cite[\S 14]{Souriau70}, the evolution space gains a unitary vector $\bu \in S^2$ such that $y = (\bx, \bv, s, \bu) \in V$, with the Carrollean action $a_V(y) = (A \bx + \bc, A \bv + \bb, s - \overline{\bb} A \bx + f, A \bu)$, and is endowed with the presymplectic 2-form, defined as,
\begin{equation}
\label{sigma_3d_free_spin}
\sigma(\delta y, \delta' y) = m  \la \delta \bv, \delta' \bx \ra - m \la \delta' \bv, \delta \bx\ra - \spin \la \bu, \delta \bu \times \delta' \bu \ra,
\end{equation}
where $\spin$ is the scalar spin (or longitudinal spin) of the particle, such that $\bl = \bx \times \bp + \spin \bu$ is a conserved quantity along the trajectory.

The equations of motion are once again readily computed (upon using that $\bu$ is unitary),
\begin{equation}
\delta(\bx, \bv, s, \bu) \in \ker(\sigma)
\;
\Leftrightarrow
\;
\delta\bx = 0, \delta \bv = 0, \delta{s} \in \R, \delta \bu = 0.
\label{kersigma_spin}
\end{equation}
Thus, in the free case, the direction of the spin of the particle is conserved on its worldline, together with its position and velocity.

\subsubsection{Dimension d = 2}

\label{model_2d_free_spinless}

Now, as we have seen in section \ref{ss:ecarr}, in dimension $d = 2$, the Carroll group admits two non-trivial central extensions, and thus the space of motions will be this time symplectomorphic to the coadjoint orbit of the central extension of the Carroll group on the moments representing elementary particles.

Let us now build a model for a planar Carroll elementary particle represented by the moment $\mu_0 = (0, \bzero, \bzero, m, q_1, q_2)$, $m > 0$, \ie a massive spinless particle with two ``charges'' $q_1$ and $q_2$. We start by computing the Maurer-Cartan form $\Theta \in \Omega^1(\eCarr(3), \ecarr(3))$, for $a = (R, \bv, \bx, s, a_1, a_2) \in \eCarr(3)$, 
\begin{equation}
\label{maurer_cartan}
\Theta(a) = \left(R^{-1} dR, R^{-1} d\bv, R^{-1} d\bx, ds + \la \bv , d\bx \ra, da_1 - \bx \times d\bx, da_2 + \bv \times d\bv\right).
\end{equation}

The pairing \eqref{pairing_ecarr} of $\mu_0$ and the Maurer-Cartan 1-form then lead to the left-invariant 1-form,
\begin{equation}
\label{varpi_2d}
\varpi := \mu_0 \cdot \Theta(a) = \la m \bv , d\bx \ra + m ds + q_1 \left(da_1 - \bx \times d\bx\right) + q_2 \left(da_2 + \bv \times d\bv\right).
\end{equation}

The evolution space $V = \eCarr(3) / \SO(2) \ni y = (\bx, \bv, s, w, z)$ is thus endowed with the following 2-form $\sigma = d \varpi$,
\begin{equation}
\label{2form_ext}
\sigma = m d\overline{\bv} \wedge d\bx - q_1 \epsilon_{ij} dx^i \wedge dx^j + q_2 \epsilon_{ij} dv^i \wedge dv^j,
\end{equation}
and becomes the presymplectic space $(V, \sigma)$. Let us now study the kernel of $\sigma$. We readily find, 
\begin{equation}
\label{ker_sigma_ext}
\delta(\bx, \bv, s, w, z) \in \ker(\sigma)
\; \Leftrightarrow \;
\left\lbrace
\begin{array}{l}
\displaystyle m \delta \bx = - 2 q_2 \epsilon \delta \bv, \\
\displaystyle m \delta \bv = - 2 q_1 \epsilon \delta \bx, \\
\delta s \in \R, \\
\delta w \in \R, \\
\delta z \in \R. \\
\end{array}
\right.
\end{equation}

Compatibility between the first two conditions implies two cases based on the value of the \emph{effective mass} squared
\begin{equation}
\label{eff_mass_free}
\tm^2 := m^2 + 4 q_1 q_2.
\end{equation}

If $\tm^2 \neq 0$, we have $\delta \bx = \delta \bv = 0$, \ie the same trivial dynamics as for the non extended Carroll group \eqref{kersigma_spinless}, and $\sigma$ is presymplectic of rank $4$, with $\dim \ker \sigma = 3$. The space of motions $U = V / \ker(\sigma)$ thus has the same topology as for the non extended Carroll group, see section the spinless part of \ref{model_3d_free}, but it is endowed with a different symplectic form, namely,
\begin{equation}
\label{omega_2d_free}
\omega = m d\overline{\bv} \wedge d\bx - q_1 \epsilon_{ij} dx^i \wedge dx^j + q_2 \epsilon_{ij} dv^i \wedge dv^j.
\end{equation}

This symplectic form now allows for an interpretation for the two central charges $q_1$ and $q_2$. Indeed, the second charge with the term $q_2 \epsilon_{ij} dv^i \wedge dv^j$ in the symplectic form seems well established in the planar literature. It appears for instance in the planar Galilean case, see \cite{DuvalH01} where it was dubbed to be the ``exotic term'' of the 2-form, and it also appears for the planar Poincaré group \cite{Feher86}. The first charge, however, seems new. As the later coupling of the particle to an external electromagnetic field will suggest, it seems to be some kind of intrinsic magnetic field (times an electric charge). 

\medskip

If the effective mass $\tm^2$ vanishes, then the first two conditions of \eqref{ker_sigma_ext} degenerate and we are left with only $m \delta \bx = - 2 q_2 \epsilon \delta \bv$, and $\dim \ker \sigma = 5$. 

\medskip 

Let us quickly mention the conserved quantities of this dynamical system. The extended Carroll group acts on $V$, with $a = \left(A, \bb, \bc, f, a_1, a_2\right) \in \eCarr(3)$, through,
\begin{equation}
\label{action_ecarr_v}
a_V
\left(\begin{matrix}
\bx \\
\bv \\
s \\
w \\
z 
\end{matrix}\right)
=
\left(\begin{matrix}
A \bx + \bc \\
A \bv + \bb \\
s + f - \la \bb, R\bx\ra \\
w + a_1 - \la \epsilon \bc, R \bx \ra \\
z + a_2 - \la \bb, R \epsilon \bv \ra
\end{matrix}\right),
\end{equation}
which leads to a representation of the Lie algebra $\ecarr(3)$ \eqref{alg_carr_ext} on vector fields $Z_V \in T_y V$,
\begin{equation}
\label{repr_vect_ecarr}
Z_V = \left(\overline{j(\omega) \bx} + \overline{\bgamma}\right) \partial_\bx + \left(\overline{j(\omega) \bv} + \overline{\bbeta}\right) \partial_\bv + \left(- \la \bbeta, \bx \ra + \varphi \right) \partial_s + \left(\alpha_1 + \la \bgamma, \epsilon \bx\ra \right) \partial_w + \left(\alpha_2 - \la\bbeta, \epsilon \bv\ra\right) \partial_z.
\end{equation}

This representation on vector fields now permits the use of Souriau's moment map \cite[\S 12]{Souriau70} $J : V \rightarrow \ecarr(3)^*$ to find conserved quantities on the worldlines of the elementary particles described by the model \eqref{2form_ext}, defined by
\begin{equation}
\sigma(Z_V) = -d (J \cdot Z), \forall Z \in \ecarr(3).
\end{equation}

Writing $J = (l, \bg, \bp, m, q_1, q_2)$, we find the following conserved quantities respectively associated to rotations, boosts, translations, and the 3 generators in the center of the group,
\begin{subequations}
\begin{empheq}[left=\empheqlbrace]{align}
l & = m \bv \times \bx + q_1 \bx^2 - q_2 \bv^2 + \theta, \\
\bg & = m \bx + 2 q_2 \epsilon \bv, \\
\bp & = m \bv + 2 q_1 \epsilon \bx, \\
m, & \\
q_1, & \\
q_2, &
\end{empheq}
\end{subequations}
for some constant $\theta$. Now, to interpret this constant, let us plug the above conserved quantities into the expression of the second Casimir invariant $C_2$ \eqref{c2} related to the anyonic spin. We easily find $C_2 = \left(1 + 4\frac{q_1 q_2}{m^2}\right) \theta$, \ie that the constant $\theta$ is (up to renormalization by some Casimir invariants) the anyonic spin of the particle.

Hence, if we wish to study spinless particles, we should set $\theta = 0$. However, as we see here, without external fields, the dynamics of a Carrollian planar anyon do not differ from those of planar particles without spin. The spin is only described by one number in the plane, and it is an invariant. 

\subsection{Dynamics of Carroll particles in an electromagnetic field}

The Maxwell-Carroll equations for electromagnetism have been derived in \cite{DuvalGHZ14,HenneauxS21,BoerHOSV21}. We are thus going to study the dynamics of massive and charged Carroll particles in an electromagnetic field, in $3+1$ and $2+1$ dimensions. 

\subsubsection{In 3+1 dimensions}
\label{model_3d_em}

To obtain the model describing the motions of massive Carroll elementary particles of charge $q$ in an electromagnetic field $F$, we are going to use the minimal coupling procedure from \cite[\S 15]{Souriau70} on the spinless free model from \ref{model_3d_free}, \ie the Carrollean limit of $\sigma \rightarrow \sigma + q F$.

The evolution space is still given by $V = \Carr(3+1)/\SO(3) \ni y = (\bx, \bv, s)$, but it is now endowed with the presymplectic 2-form,
\begin{equation}
\label{sigma_3d_em_spinless}
\sigma = \left(m d \overline{\bv} - q \overline{\bE} ds\right) \wedge d\bx + \half q B^i \epsilon_{ijk} dx^j \wedge dx^k,
\end{equation}
where we finally see the introduction of the Carrollean time in the 2-form with the help of the electric field. Note that the electric field has dimensions such that $[q E/m] = L^{-1}$ due to some dimensional rescaling of the electromagnetic fields.

We then readily characterize the kernel of $\sigma$,
\begin{equation}
\label{ker_sigma_3d_em_spinless}
\delta(\bx, \bv, s) \in \ker(\sigma) \; \Leftrightarrow \; \delta \bx = 0, \delta \bv = \frac{q}{m} \bE \delta s, \delta s \in \R, \delta w \in \R, \delta z \in \R.
\end{equation}

In other words, we have the equations of motion,
\begin{subequations}
\label{eom_3d_em_spinless}
\begin{align}
\frac{d \bx}{ds} & = 0, \\
\frac{d (m \bv)}{ds} & = q \bE.
\end{align}
\end{subequations}
We notice here the extreme decoupling in Carroll dynamics of the momentum and the velocity of the particle. The particle still does not move, even in an electromagnetic field, however its momentum feels the electric field. Recall that the energy is \emph{not} a conserved quantity in Carroll dynamics. The magnetic field is transparent to the Carroll particle, which is to be expected since the particle does not move. 

\medskip
Let us now consider the model of a massive elementary particle with spin \eqref{sigma_3d_free_spin}, \ie described by the moment $\mu_0 = (s, 0, 0, m)$, to which we add the Carrollean limit of the spin-magnetic field coupling term \cite[\S 15]{Souriau70}. We thus have the evolution space $V \ni y = (\bx, \bv, s, \bu)$ endowed with the presymplectic 2-form,
\begin{equation}
\label{sigma_3d_em_spin}
\begin{array}{l}
\sigma(\delta y, \delta' y) = m  \la \delta \bv - q \bE \delta s, \delta' \bx \ra - m \la \delta' \bv - q \bE \delta s', \delta \bx\ra - \spin \la \bu, \delta \bu \times \delta' \bu \ra \\
\qquad\qquad\qquad + q \la \bB, \delta \bx \times \delta ' \bx\ra + \mu \left(\delta (\la \bu, \bB \ra) \delta' s - \delta' (\la \bu, \bB \ra) \delta s\right),
\end{array}
\end{equation}
where $\mu$ is the magnetic moment of the particle (not to be confused with the moment map). The equations of motion are then,
\begin{subequations}
\label{eom_3d_em_spin}
\begin{align}
\frac{d \bx}{ds} & = 0, \\
\frac{d (m \bv)}{ds} & = q \bE + \mu \bnabla \la \bu, \bB \ra, \label{eom_3d_em_spin_dv}\\
\frac{d \bu}{ds} & = \mu \, \bu \times \bB. \label{eom_3d_em_spin_du}
\end{align}
\end{subequations}

The equations are similar to those from the Galilean case \cite[\S 15]{Souriau70}. The differences being the substitution $t \rightarrow s$ and the vanishing Carrollean velocity $d\bx / ds = 0$. These equations show a precession of the spin around the magnetic field.

\subsubsection{In 2+1 dimensions}
\label{model_2d_em}

Let us now consider dynamics in $2+1$ dimensions, where things get more interesting, by presenting some actual motions.

In the same spirit as the previous section, we apply the minimal coupling of Carroll electromagnetism to a planar Carrollean elementary particle from the free model \eqref{model_2d_free_spinless}. However, we will consider particles with spin, \ie anyons, from the start. Also, recall that, due to the electromagnetic tensor $F$ being skewsymmetric, in $2+1$ dimensions, the electric field has 2 components, while the magnetic field has only 1 component. We readily get the presymplectic 2-form on the evolution space $V \cong \eCarr(3) / \SO(2) \ni y = (\bx, \bv, s, w, z)$,
\begin{equation}
\label{sigma_2d_em}
\sigma = \left(m d \overline{\bv} - q \overline{\bE} ds\right) \wedge d\bx - \left(q_1 - \half q B\right)  \epsilon_{ij} dx^i \wedge dx^j + q_2 \epsilon_{ij} dv^i \wedge dv^j + \mu \, \theta \, dB \wedge ds,
\end{equation}
where $B$ is seen as a function of the spacetime coordinates $\bx$ and $s$, and $\mu$ is the coupling constant between the anyon of spin $\theta$ and the magnetic field.

The kernel of the 2-form is now characterized by,
\begin{equation}
\label{ker_sigma_2d_em}
\delta(\bx, \bv, s, w, z) \in \ker(\sigma)
\; \Leftrightarrow \;
\left\lbrace
\begin{array}{l}
q \la \bE, \delta \bx \ra = - \mu \, \theta \la \bnabla B, \delta \bx \ra, \\
m \delta \bx = -2 q_2 \epsilon \delta \bv, \\
m \delta \bv = q \bE \delta s - \left(2 q_1 - q B\right) \epsilon \delta \bx + \mu \, \theta \, \bnabla B \, \delta s, \\
\delta s \in \R, \\
\delta w \in \R, \\
\delta z \in \R
\end{array}
\right.
\end{equation}
which leads to the equations of motion, if the \emph{effective mass} squared 
\begin{equation}
\label{eff_mass_em}
\tm^2 := m^2 + 4 \left(q_1 - \half q B\right) q_2,
\end{equation}
(which reduces to the effective mass define in \eqref{eff_mass_free} if we turn off the magnetic field) does not vanish,
\begin{subequations}
\label{eom_2d_em}
\begin{align}
\frac{d\bx}{ds} & = - \frac{2 q_2}{\tm^2} \epsilon \left(q \bE+ \mu \theta \bnabla B\right), \\
\frac{d(m \bv)}{ds} & = \frac{m^2}{\tm^2} \left(q \bE + \mu \theta \bnabla B\right).
\end{align}
\end{subequations}

If the effective mass vanishes, the first and third equations characterizing the kernel \eqref{ker_sigma_2d_em} yield $m \delta \bx = - q_2 \epsilon \delta \bv$ and $\left(q \bE + \mu \theta \bnabla B\right) \delta s = 0$. Hence, the equations of motion become effectively the same as in the $3+1$ dimensional case where the mass vanishes.

In planar Galilean dynamics, it was shown that the limiting case of effective mass vanishing corresponds to some kind of Hall motions \cite{DuvalH01}. It would thus be interesting to investigate this situation further in the Carrollean case. It is, however, rendered much more complicated than its Galilean cousin since the Carrollean velocity is not related to the Carrollean momentum.

Finally, we recover the non extended model equations of motion \eqref{eom_3d_em_spinless} when $q_1 = q_2 = 0$, and those of the free model \ref{model_2d_free_spinless} if instead we turn off both electromagnetic fields.

\bigskip

Let us now consider the important particular case where the Casimir invariant $C_1$ that we call the mass vanishes, \ie $m = 0$, and with electric charge $q = 0$, which should describe a planar Carrollean photon. Such particles were first mentioned in \cite{DuvalGH14b}, however without central extensions. We suppose that such particle would still have both ``exotic'' charges $q_1$ and $q_2$, and anyon spin $\theta$. Hence, this case happens when the presymplectic form \eqref{ker_sigma_2d_em} reduces to,
\begin{equation}
\label{sigma_2d_em_photon}
\sigma = - q_1 \epsilon_{ij} dx^i \wedge dx^j + q_2 \epsilon_{ij} dv^i \wedge dv^j + \mu \, \theta \, dB \wedge ds,
\end{equation}

The equations of motion are easily obtained, and read,
\begin{subequations}
\label{eom_2d_em_photon}
\begin{align}
\frac{d\bx}{ds} & = - \frac{\mu \theta}{2 q_1} \epsilon \bnabla B, \\
\frac{d(m \bv)}{ds} & = 0.
\end{align}
\end{subequations}

We notice that they can be obtained from the previously computed equations of motion \eqref{eom_2d_em} by setting $m = 0$ and $q = 0$. Hence, the massless limit from the massive equations of motion is regular. 

\medskip

It is quite interesting to note that in the massive case \eqref{eom_2d_em} the main ``coupling constant'' that brings motion is the second charge $q_2$, while in the massless case \eqref{eom_2d_em_photon} this rôle is played by $q_1$. 

\subsection{Dynamics of Carroll particles in a gravitational field}
\label{ss:grav}

In this section, we will exclusively focus on the $2+1$ dimensional case, as the dynamics of Carroll particles in $3+1$ dimensions and higher have already been studied in \cite{BergshoeffGL14}. However, some of our results will be valid for any dimensions, and we will recover their results in one configuration.

While the coupling of a particle to an electromagnetic field is rather straightforward, in that it requires an additional interaction term in the presymplectic 2-form, see \eg \eqref{sigma_3d_em_spinless}, the coupling to a gravitational field is more subtle as the interaction is modeled through a curved background. Such a presymplectic construction in a curved background is well understood for Lorentzian mechanics, see \cite{Kun72}, and for Galilean/Bargmannian system in a (Newtonian) gravitational field, see \cite{TheseChristian}. The construction of the extended Carroll symplectic geometry follows the same logic as the Galilean one due to the presence of the central extension.

\subsubsection{Geometric framework}

The idea of the construction is rather simple. The first step is to notice that in the free case the potential 1-form \eqref{varpi_2d} of the presymplectic 2-form is defined in terms of the Maurer-Cartan form $\Theta$ of a group $G$ as $\varpi = \mu_0 \cdot \Theta$ for some moment $\mu_0$ in the orbit that we wish to describe. Thus, we can see this free potential 1-form as living on the flat geometry defined by the Klein pair $(G, H)$, for some $G$ and $H$, naturally equipped with the Maurer-Cartan form as a flat connection. The generalization to a curved space is then natural: by considering the Cartan geometry based on the Klein model $(G, H)$, and by replacing the Maurer-Cartan form in the definition of the presymplectic potential by a Cartan connection. 

In order to do this in the Carrollean case, we are going to consider two Cartan geometries. The first one is the based on the Klein pair $(G, H)$ where $G$ is the Carroll group and $H$ its subgroup without translations, which we will realized inside the Carroll frame bundle $H(M)$. The second Cartan geometry is the ``extended'' Cartan Carroll geometry, this time with $G$ the extended Carroll group, and $H$ again its subgroup without translations. In both cases, the base manifold $G/H = M$ is the Carrollean spacetime. The first geometry is the ``physical'' one, while the second one is a mathematical tool in order to define a curved presymplectic 2-form. We wish that in the end, everything projects down on the Carroll frame bundle.

\medskip

Let us first consider the Carroll frame bundle $H(M)$ above spacetime $M$ for a Carroll structure $(M, g, \xi)$, with local coordinates $(x^\mu, e^\mu{}_a)$. The tetrad $(e^\mu{}_a)$ is linked to the Carroll metric through,
\begin{subequations}
\begin{align}
\delta_{AB} & = g_{\mu\nu} e^\mu{}_A e^\nu{}_B, \\
\xi^\mu & = e^\mu{}_0, \label{cartan_xi}
\end{align}
\end{subequations}
where $\mu, \nu = 0, 1, 2$ are spacetime indices, $A, B = 1, 2$ are form indices that run over space, and $a, b = 0, 1, 2$ are form indices that run over space and time. Note that since the metric is degenerate, we need a second relation to completely define the tetrad, which is done with the ``time'' vector field $\xi$ in \eqref{cartan_xi}. In the end, the tetrad itself is well defined and invertible. 

Recall that the soldering form $\theta$ and the most general linear connection $\womega^a{}_b$ on a frame bundle of coordinates $(x^\mu, e^\mu{}_a)$ are given by,
\begin{subequations}
\label{cartan_gl}
\begin{align}
\theta^a & = \theta^a{}_\mu dx^\mu, \\
\womega^a{}_b & = \theta^a{}_\mu\left(d e^\mu{}_b + \Gamma^\mu_{\nu\lambda} e^\nu{}_b dx^\lambda\right),
\end{align}
\end{subequations}
for some functions $\Gamma^\mu_{\nu\lambda}$, and where we have $\theta^a{}_\mu e^\mu{}_b = \delta^a_b$ and $e^\mu{}_c \theta^c{}_\nu = \delta^\mu_\nu$.

Given the representation \eqref{alg_mat_carroll} of the Carroll Lie algebra, a Cartan connection $\womega$ takes the form,
\begin{equation}
\womega = \left(\begin{matrix}
\womega^A{}_B & 0 & \theta^A \\
- \womega^0{}_B & 0 & \theta^0 \\
0 & 0 & 0
\end{matrix}\right).
\end{equation}

Note that in the case of a Carroll Cartan geometry, we have $\womega^A{}_0 = 0$ and $\womega^0{}_0 = 0$. 

We define the curvature 2-form $\Omega = d\womega + \womega \wedge \womega$ through the usual structure equations,
\begin{subequations}
\begin{align}
\Omega^A{}_B & = d \womega^A{}_B + \womega^A{}_C \wedge \womega^C{}_B, \\
\Omega^4{}_B & = d \womega^0{}_B + \womega^0{}_C \wedge \womega^C{}_B, \\
\Omega^A & = d \theta^A + \womega^A{}_C \wedge \theta^C, \\
\Omega^4 & = d \theta^0 - \womega^0{}_C \wedge \theta^C,
\end{align}
\end{subequations}
where we have the usual definitions \cite{KobayashiN63} of the torsion tensor $T$ through $\Omega^a = \half T^a_{bc} \theta^b \wedge \theta^c$, and $\Omega^a{}_b = \half R^a{}_{bcd} \theta^c \wedge \theta^d$ with $R$ the Riemann tensor.

\bigskip

Consider now the Cartan geometry of the extended Carroll algebra. 
Given the representation \eqref{alg_mat_ecarr} of the algebra, a Cartan connection is then parametrized as,
\begin{equation}
\womega = \left(\begin{matrix}
\womega^A{}_B & 0 & \theta^A & 0 & \epsilon^A{}_C \delta^{CD} \womega^0{}_D \\
- \womega^0{}_B & 0 & \theta^0 & 0 & \womega_2 \\
0 & 0 & 0 & 0 & 0 \\
\theta^C \delta_{CD} \epsilon^D{}_B & 0 & \womega_1 & 0 & - \theta^0 \\
0 & 0 & 0 & 0 & 0
\end{matrix}\right),
\end{equation}
with structure equations,
\begin{subequations}
\label{structure_eq_ecarr}
\begin{align}
\Omega^A{}_B & = d \womega^A{}_B + \womega^A{}_C \wedge \womega^C{}_B, \\
\Omega^0{}_B & = d \womega^0{}_B + \womega^0{}_C \wedge \womega^C{}_B, \\
\Omega^A & = d \theta^A + \womega^A{}_C \wedge \theta^C, \\
\Omega^0 & = d \theta^0 - \womega^0{}_C \wedge \theta^C, \\
\Omega_1 & = d\womega_1 + \theta^C \wedge \theta^D \epsilon_{CD}, \\
\Omega_2 & = d\womega_2 - \womega^0{}_C \wedge \womega^0_D \epsilon^{CD},
\end{align}
\end{subequations}
where $\Omega_1$ and $\Omega_2$ are the curvature terms associated to the central extension parameters. From now on, for the sack of simplicity, we will ask for the torsion to vanish: $\Omega^a = 0$.

One may now define the symplectic potential associated to this geometry as $\varpi = \mu_0 \cdot \womega$. Considering a spinless massive particle, with the same parameters as in \ref{model_2d_free_spinless}, this potential 1-form is,
\begin{equation}
\label{varpi_cartan}
\varpi = m \theta^0 + q_1 \womega_1 + q_2 \womega_2.
\end{equation}
Note that, due to the second and third terms in the above expression, the presymplectic potential does not project down to the Carroll frame bundle.

The presymplectic 2-form is then $\sigma = d \varpi = m d \theta^0 + q_1 d\womega_1 + q_2 d\womega_2$ or, upon using the structure equations \eqref{structure_eq_ecarr},
\begin{equation}
\label{sigma_2d_curved}
\sigma = m \womega^0{}_C \wedge \theta^C - q_1 \theta^C \wedge \theta^D \epsilon_{CD} + q_1 \Omega_1 + q_2 \womega^0{}_C \wedge \womega^0{}_D \epsilon^{CD} + q_2 \Omega_2.
\end{equation}

Since $\Omega_1$ and $\Omega_2$ are curvature terms, they are tensorial forms and may be written as $\Omega_1 := \half  \Omega_{1,ab} \theta^a \wedge \theta^b$ and $\Omega_2 := \half \Omega_{2,ab} \theta^a \wedge \theta^b$. Thus, the presymplectic 2-form \eqref{sigma_2d_curved} turns out to project down to the Carroll frame bundle $H(m)$, even though the potential 1-form does not. 

\subsubsection{Equations of motion without exotic curvature terms}

Let us now obtain the equations of motion of the above model when both extension curvature terms vanish. We have then the 2-form,
\begin{equation}
\label{sigma_2d_curved_noomega}
\sigma = m \womega^0{}_C \wedge \theta^C - q_1 \theta^C \wedge \theta^D \epsilon_{CD} + q_2 \womega^0{}_C \wedge \womega^0{}_D \epsilon^{CD}.
\end{equation}

Notice the similarity between the expression of this 2-form on $H(M)$ and \eqref{2form_ext}. When comparing the potential form \eqref{varpi_cartan} on this Cartan geometry with the flat case of \ref{model_2d_free_spinless}, we see that one should recover $\theta^0 = ds + \overline{\bv} d\bx$ in the flat case.

To study the kernel of such a presymplectic 2-form, let us define a vector field $X \in T_{(x,e)}(H(M))$,
\begin{equation}
\label{champ_X}
X = \frac{dx^\mu}{d\tau} \partial_\mu + \frac{d e^\mu{}_a}{d\tau} \partial_{e^\mu{}_a},
\end{equation}
for some parameter $\tau$. Define then
\begin{subequations}
\label{dotxe}
\begin{align}
\dot{x}^\mu & := \frac{dx^\mu}{d\tau}, \\
\dot{e}^\mu{}_a & := \frac{de^\mu{}_a}{d\tau} + \Gamma^\mu_{\nu\lambda} \dot{x}^\nu e^\lambda{}_a.
\end{align}
\end{subequations}

Combining \eqref{cartan_gl}, \eqref{champ_X}, and \eqref{dotxe}, we find $\theta^a(X) = \theta^a{}_\mu \dot{x}^\mu$ and $\womega^a{}_b(X) = \theta^a{}_\mu \dot{e}^\mu{}_b$, which imply,
\begin{equation}
\sigma(X) = m \theta^0{}_\mu \dot{e}^\mu{}_C \theta^C - m \theta^C{}_\mu \dot{x}^\mu \womega^0{}_C - 2 q_1 \theta^B{}_\mu \dot{x}^\mu \theta^C \epsilon_{BC} + 2 q_2 \theta^0{}_\mu \dot{e}^\mu{}_C \womega^0{}_D \epsilon^{CD}.
\end{equation}

The equations of motion are characterized by vector fields $X$ which annihilate the above expression. We thus find two conditions that characterize the kernel of $\sigma$, $m \theta^0{}_\mu \dot{e}^\mu{}_C - 2 q_1 \theta^B{}_\mu \dot{x}^\mu \epsilon_{BC} = 0$ and $m \theta^C{}_\mu \dot{x}^\mu - 2 q_2 \theta^0{}_\mu \dot{e}^\mu{}_B \epsilon^{BC} = 0$. These conditions lead to,
\begin{subequations}
\begin{align}
\left(m^2 + 4 q_1 q_2\right) \theta^A{}_\mu \dot{x}^\mu & = 0, \\
\left(m^2 + 4 q_1 q_2\right) \dot{\theta}^0{}_\mu e^\mu{}_A & = 0.
\end{align}
\end{subequations}

The discussion is essentially the same as for the free case in \ref{model_2d_free_spinless}, but in a slightly different formalism: if the effective mass $\tm^2 = m^2 + 4 q_1 q_2$ does not vanish, then we find that $\dot{x}^\mu \propto e^\mu{}_0$ and $\dot{\theta}^0{}_\mu \propto \theta^0{}_\mu$. Given that $\womega^0{}_0 = 0$, we have $\womega^0{}_0(X) = \dot{\theta}^0{}_\mu e^\mu{}_0 = 0$, meaning that we also have $\dot{\theta}^0{}_\mu \propto \theta^A{}_\mu$. Since $\theta^A{}_\mu$ and $\theta^0{}_\mu$ are independent, we find $\dot{\theta}^0{}_\mu = 0$. Lastly, given the definitions \eqref{cartan_xi} and $\theta^0{}_\mu := v_\mu$ (for that last definition, see the comparison with the flat case after \eqref{sigma_2d_curved_noomega}), we find,
\begin{subequations}
\begin{align}
\dot{x}^\mu & = \lambda \xi^\mu, \label{eom_curved_dx} \\
\dot{v}_\mu & = 0,
\end{align}
\end{subequations}
for $\lambda \in \R^*$. These motions are formally the same as those obtained in \cite{BergshoeffGL14} when considering the gravitational coupling for spatial dimension $d \geq 3$: the Carrollean velocity is along the direction defined by $\xi$. 

Recall that the vector field $\xi$ is nowhere vanishing by definition on a Carroll structure. It is thus natural to take this vector field as the definition of the time direction, and denote $\xi = \partial_s$, such that $\xi^0 = 1 = e^0{}_0$ and $\xi^A = 0$ with $x^0 = s$. In that case, upon choosing an appropriate parameter $\lambda \neq 0$, we find from \eqref{eom_curved_dx}, $ds/d\tau = \lambda$, and thus,
\begin{subequations}
\begin{align}
\frac{d \bx}{ds} & = 0, \\
\frac{D \bv}{ds} & = 0,
\end{align}
\end{subequations}
where $D/ds$ is the covariant derivative. 

\medskip

We could have obtained the above equations of motion directly from the flat equations of motion \eqref{eom_free} by applying the ``minimal coupling'' procedure, \ie by replacing the derivative on the momentum by a covariant derivative. Given that in Carrollean dynamics the Carrollean momentum is completely decoupled from the Carrollean velocity, it is clear that minimal coupling to gravity should have no impact on dynamics. The two central extensions thus have no impact on the dynamics of Carroll particles in a gravitational field.

\subsubsection{Equations of motion with exotic curvature terms}

Let us now finish by considering the general case of the curved presymplectic space where one has non vanishing curvature forms $\Omega_1$ and $\Omega_2$ associated to the central extensions. Note that these terms, while being curvature terms, are not linked to the Riemann tensor, and thus do not represent gravitational coupling. Since we consider the 2+1 dimensional case, one may decompose these 2-forms into a ``magnetic'' part and an ``electric'' part, just like for the electromagnetic tensor, as $\Omega_1 := \half \Omega_{1,ab} \theta^a \wedge \theta^b := \cO_1 \epsilon_{AB} \theta^A \wedge \theta^B + \Omega_{1,B} \theta^0 \wedge \theta^B$, and similarly for $\Omega_2$.

It is immediate to see that the magnetic term of these curvatures do not play any rôle in the dynamics if they are present without the electric terms. Indeed, from their decomposition and the presymplectic 2-form \eqref{sigma_2d_curved}, we see that the consideration of these terms is equivalent to a shift on the first central extension parameter: $q_1 \mapsto q_1 (1 - \cO_1 - q_2 \cO_2 / q_1)$. Since $q_1$ is arbitrary, we can absorb those terms into its definition.

\medskip

It remains to study the electric part of these curvature terms. The presymplectic 2-form \eqref{sigma_2d_curved} becomes,
\begin{equation}
\sigma = m \womega^0{}_C \wedge \theta^C - q_1 \theta^C \wedge \theta^D \epsilon_{CD} + q_2 \womega^0{}_C \wedge \womega^0{}_D \epsilon^{CD} + T_A \theta^0 \wedge \theta^A,
\end{equation}
where $T_A := q_1 \Omega_{1,A} + q_2 \Omega_{2,A}$. Using the same procedure as in the previous section, the kernel of $\sigma$ is characterized by the following conditions,
\begin{subequations}
\begin{align}
& m \theta^0{}_\mu \dot{e}^\mu{}_C - 2 q_1 \theta^B{}_\mu \dot{x}^\mu \epsilon_{BC} + T_A \theta^0{}_\mu \dot{x}^\mu = 0, \\
& T_A \theta^A{}_\mu \dot{x}^\mu = 0, \\
& m \theta^C{}_\mu \dot{x}^\mu - 2 q_2 \theta^0{}_\mu \dot{e}^\mu{}_B \epsilon^{BC} = 0.
\end{align}
\end{subequations}

From the second and third equations, we find $\theta^A{}_\mu \dot{x}^\mu = \lambda \epsilon^{AB} T_B$ and $\dot{\theta}^0{}_\mu e^\mu{}_B = \frac{m \lambda}{2 q_2} T_B$ for some $\lambda \in \R^*$. By injecting these two relations into the first, we find $\theta^0{}_\mu \dot{x}^\mu = \lambda \frac{\widetilde{m}^2}{2 q_2}$, where we have the usual modified mass $\widetilde{m}^2 = m^2 + 4 q_1 q_2$, and we always have, due to the Carroll algebra, $\dot{\theta}^0{}_\mu e^\mu{}_0 = 0$. All these relations together with the definitions $\xi^\mu = e^\mu{}_0$, and $\theta^0{}_\mu = v_\mu$ lead to the equations of motion, after absorbing the factor $\lambda \neq 0$ into the parameter $\tau$,
\begin{subequations}
\begin{align}
\dot{x}^\mu & = \frac{\widetilde{m}^2}{2 q_2} \xi^\mu + e^\mu{}_A \epsilon^{AB} T_B, \\
\dot{v}_\mu & = \frac{m}{2 q_2} T_A \theta^A{}_\mu.
\end{align}
\end{subequations}

We find actual motions with a non trivial spatial component for the velocity. Note that the equations of motion we obtain in this gravitational background are formally the same as those we obtained earlier in an electric background \eqref{eom_2d_em} (without the magnetic term). Indeed, we find that the spatial velocity is orthogonal to the ``electric'' vector $T_A$, while the derivative of the momentum is in the same direction as $T_A$.

An interesting question pops up, however: what is the physical meaning of $T_A = q_1 \Omega_{1,A} + q_2 \Omega_{2,A}$? While $q_1$ and $q_2$ are properties of the elementary particle, the $\Omega_{1,A}$ and $\Omega_{2,A}$ are source terms, but not gravitational sources. Their physical interpretation thus remains open. 

\section{Carroll quantum equations}
\label{s:quantum}

Much like how the Klein-Gordon equation results from the Poincaré group, and the Schr\"odinger equation results from the Galilei group, one can wonder what is the quantum equation associated to the Carroll group. This can be answered in different ways.

\subsection{Casimir considerations}

First, the quantum equation can be intuited from the Casimir invariants of the group. Recall that one of the Casimir invariants of the Poincaré group is $p_0{}^2 = \bp^2 + m^2$, which leads, under the prescription that $p_0 \rightarrow i \hbar \partial_t$ and $\bp \rightarrow - i \hbar \nabla$, to the free Klein-Gordon equation. For the Galilei group, one of the Casimir invariants is $p_0 = \frac{\bp^2}{2m}$, which leads to the free Schr\"odinger equation under this same prescription. Now, for the Carroll group, the first Casimir of the group is $p_0 = m$. We can thus expect the free quantum equation to be of the form,
\begin{equation}
\label{carroll_quant_1}
- i \hbar \partial_s \psi(\bx, s) = m \psi(\bx, s).
\end{equation}

\subsection{Carrollean limit of the Klein-Gordon equation}

As a second approach, consider the free Klein-Gordon equation, which reads in coordinates $(\bx, x^4)$,
\begin{equation}
\left(\Delta - (\partial_0)^2 - \frac{m^2c^2}{\hbar^2}\right) \psi(\bx, x^0) = 0
\end{equation}
where $\Delta$ is the Laplacian. It is well known that if we define $x^0 := c t$, the limit $c \rightarrow \infty$ of the above equation yields, after the redefinition $\psi(\bx, t) \rightarrow \psi(\bx, t) \exp(-i m c^2 t/\hbar)$, the usual (free) Schr\"odinger equation for $\psi$. Now, let us apply the Carroll limit by renaming the velocity as $C$ and defining the time coordinate as $x^0 := s/C$. The Klein-Gordon equation is then $\left(\frac{1}{C^2}\Delta - (\partial_s)^2 - \frac{m^2}{\hbar^2}\right) \psi(\bx, s) = 0$. After taking the limit $C \rightarrow \infty$, we immediately get,
\begin{equation}
\label{carroll_quant_2}
\left(\partial_s\right)^2 \psi(\bx, s) = - \frac{m^2}{\hbar^2} \psi(\bx, s),
\end{equation}
which is the ``square'' of the equation \eqref{carroll_quant_1} obtained from the Casimir invariant. 

\subsection{Geometric quantization for massive and spinless Carroll particles, \texorpdfstring{$d \geq 3$}{d >= 3}}

As a third, and final, approach, we are going to apply geometric quantization \cite{Kostant69,Souriau70} to the classical symplectic model defined in \ref{model_3d_free} for a free massive spinless particle.\footnote{Some of the computations in this section were done with the late Christian Duval during the author's 1st-Master year internship under his supervision in 2016.}

First, we want to construct a \emph{prequantum bundle} $(Y, \alpha)$ \cite[\S 18]{Souriau70} over the space of motions $(U, \omega)$ defined for the model of a Carrollean free spinless particle in $3+1$ dimensions in section \ref{model_3d_free}. Recall that such a prequantum bundle is a principal circle-bundle $\pi : Y \rightarrow U$ endowed with a $U(1)$-invariant 1-form $\alpha$, such that $d\alpha = \pi^* \omega$.

Now, $(U, \omega)$ is a symplectic manifold admitting a potential, hence we apply the prequantization procedure as shown in \cite[\S 18]{Souriau70}, which simply consists of defining $Y = U \times S^1 \ni y = (\bx, \bp, z)$, on which $U(1)$ acts as $z'_Y (\bx, \bp, z) = (\bx, \bp, z' z)$, and the 1-form,
\begin{equation}
\label{alpha}
\alpha = \frac{1}{\hbar} \la \bp, d\bx \ra + \frac{dz}{iz}.
\end{equation}

Note that the prequantum bundle can also be defined directly from the evolution space $V$ as the quotient $Y = V / (2\pi \hbar/m) \Z$ with $(\bx, \bp, s) \mapsto (\bx, \bp, z=e^{ims/\hbar})$, so that we have $\varpi = \hbar (V \rightarrow Y)^*\alpha$, with $\varpi$ defined in \eqref{varpi_3d}. Since $U$ is simply-connected, this prequantization is unique.

\smallskip

The pre-Hilbert space $\cH_Y$ is the set of $U(1)$-equivariant differentiable functions $\Psi : Y \rightarrow \C$ with compact support, endowed with the scalar product $\la \Phi, \Psi \ra := \int_M \overline{\Phi(y)} \Psi(y) \Omega, \, \forall \Phi, \Psi \in \cH_Y$, where $\Omega$ is the Liouville volume form of $(M, \omega)$, and where the norm is defined as $\Vert \Psi \Vert = \sqrt{\la \Psi, \Psi \ra}$ \cite[\S 18]{Souriau70}. In our case, these functions are thus of the form $\Psi(\bx, \bp, z) = z \phi(\bx, \bp)$ for some complex-valued function $\phi : U \rightarrow \C$. 

Now, Geometric Quantization requires that a polarization, which is a maximal isotropic foliation of the symplectic base manifold, see \eg \cite{Souriau70}, be chosen in order to lead to an irreducible representation of the group. In practice, this means choosing either a position or a momentum representation. Here, given the canonical form of the symplectic 2-form on $U$, $\omega = d\overline{\bp} \wedge d\bx$, we clearly have at our disposal the position polarization $\cF_x \subset TU$ generated by the distribution $\la \partial_{p_1}, \ldots, \partial_{p_d} \ra$. 

We then construct the Hilbert space $\cH_Y^{\cF_x}$ as the subset of (the completion of) $\cH_Y$ such that functions are constant along the directions of the horizontal lift $\widetilde{\cF}_x$ of the distribution $\cF_x$, \ie $\widetilde{X} \Psi = 0, \, \forall \Psi \in \cH^{\cF_x}_Y, \, \forall \widetilde{X} \in \widetilde{\cF}_x$. The quantum wave functions of the model thus consist of the functions in $\cH^{\cF_x}_Y$, which are of the form,
\begin{equation}
\label{psi_position}
\Psi_x(\bx, \bp, s) = e^{ims/\hbar} \phi_x(\bx),
\end{equation}
where $\phi \in C^\infty_c(\R^d, \C)$.

Note that we could have chosen the momentum representation by using the horizontally lifted polarization $\widetilde{\cF_p} \subset TY$ generated by the distribution $\la \partial_i - \frac{p^i}{m} \partial_s \ra_{i = 1, \ldots, d}$. This leads to wave functions of the form,
\begin{equation}
\label{psi_momentum}
\Psi_p(\bx, \bp, s) = e^{ims/\hbar} e^{i \la \bp, \bx \ra / \hbar} \phi_p(\bp).
\end{equation}

The Carrollean wave function \eqref{psi_position} is the general solution to the quantum equation we intuited in \eqref{carroll_quant_1} (and it is also a solution of \eqref{carroll_quant_2}). We can rewrite this equation in a more covariant way,
\begin{equation}
\label{carroll_quant_3}
\frac{\hbar}{i} L_\xi \Psi = m \Psi,
\end{equation}
where $L_\xi$ is the Lie derivative along the vector field $\xi = \partial_s$. This equation appears to be the only quantum condition for a Carrollean wave function $\Psi$ describing quantum spinless particles of mass $m$. This equation plays the same rôle in the Carrollean framework as the Schr\"odinger equation does in the Galilean one. Carroll wave functions are essentially defined by arbitrary functions of space.

This equation is certainly coherent with classical motions, as the evolution through ``time'' is simply described by a change of the phase factor. Hence, the probability density computed from the wave function is constant at any space point.

The equation \eqref{carroll_quant_3} is clearly invariant under $\Carr(d+1)$. However, it is not the most general group of symmetries of this equation. Recall that a general Carroll structure $(M, g, \xi)$ can be seen as a $\R$-principal bundle, where the base $Q = M / \R \xi$ is absolute space\footnote{This contrasts with a Newton-Cartan structure which is seen as a fibration over absolute time.}. The quantum equation \eqref{carroll_quant_3} is then invariant by the full group of automorphisms of $M \rightarrow Q$. It is not surprising that the group of symmetries of a quantum equation is larger than the classical group it comes from, the same happens with the Schr\"odinger equation, for instance, which is invariant under the Schr\"odinger group \cite{Niederer72}, though the group here is infinite dimensional.

For a general orientable Carroll structure, we propose to take the equation \eqref{carroll_quant_3} as the quantum Carroll equation for a particle of mass $m$ and spin zero. We will require that wave functions are square integrable over space $Q$, \ie that they are in fact half-densities of space, such that we have locally $\Psi = \psi \otimes \vert \vol_Q \vert^\half$, where $\vol_Q$ is the volume form of $Q$, and for some function $\psi : M \rightarrow \C$ of spacetime, together with,
\begin{equation}
\int_Q \vert \Psi \vert^2 < +\infty.
\end{equation}

Finally, given the general form of a Carroll wave function in the position representation \eqref{psi_position}, we can define a unitary irreducible representation of the Carroll group through $\rho(a) \Psi_\bx := \Psi_\bx \circ a_Y^{-1}, \, \forall \Psi_\bx \in \cH_Y^{\cF_x}, \, \forall a \in \Carr(d+1)$. We find, for $\Psi_\bx(\bx, \bp, s) = e^{ims/\hbar} \phi(\bx)$,
\begin{equation}
\left(\rho(a) \Psi_\bx\right)(\bx, \bp, s) = e^{i \frac{m}{\hbar}\left(s + \la \bb, \bx - \bc\ra - f\right)} \phi\left(A^{-1} \left(\bx - \bc\right)\right).
\end{equation}

Similarly, we can compute the representation of the group on wave functions in the momentum representation \eqref{psi_momentum}, leading to, for $\Psi_\bp(\bx, \bp, s) = e^{ims/\hbar} e^{i \la \bp, \bx \ra / \hbar} \phi(\bp)$,
\begin{equation}
\left(\rho(a) \Psi_\bp\right)(\bx, \bp, s) = e^{i \frac{m}{\hbar}\left(s - f\right)} e^{i \la \bp, \bx - \bc \ra / \hbar} \phi\left(A^{-1} \left(\bp - m \bb\right)\right).
\end{equation}

\section{Conclusions, interpretations and comparisons with literature}

It has been known for some time that the Carroll algebra admits two central extensions in dimension $2+1$. We expanded on these results by finding a matrix representation for this algebra~\eqref{alg_mat_ecarr}, and then, thanks to the exponential map, a representation for the twice centrally extended Carroll group \eqref{gp_mat_ecarroll}. This extended group naturally features two additional Casimir invariants $q_1$ and $q_2$, along with the mass and spin of an elementary particle, see \eqref{casimirs}.

We then went on to study the symplectic geometry of the Carroll group, both in dimensions $3+1$ and $2+1$. In the first case, we recalled that a massive Carrollean particle, with or without spin, does not move. This stays true in an electromagnetic field, as we computed in \eqref{eom_3d_em_spin}. However, for a massive particle with spin in an electromagnetic field, the equations of motion show that the direction of its spins shows a precession around the magnetic field, see \eqref{eom_3d_em_spin_du}. These dynamics emphasize a complete decoupling between the Carrollean momentum and the Carrollean velocity of such a particle.

Next, we computed the dynamics of a particle in the latter case, \ie of a Carrollean particle in $2+1$ dimensions, featuring both new ``charges'' associated to the extensions of the group. While the two charges do not play a rôle in the free case, the motions of the particles become non trivial in an electromagnetic field, leading to actual motions, as shown for massive particles with \eqref{eom_2d_em} and massless particles with \eqref{eom_2d_em_photon}. Quite interestingly, the charge $q_1$ acts as some kind of coupling constant to bring motion to massless particles, while the second charge $q_2$ does the same, but for massive particles. These charges also enter into the definition of an effective mass \eqref{eff_mass_em}, which reminds of what happens in the Galilean case, see \eg \cite{DuvalH01}. However, in the Carrollean case, unlike in the Galilean case, the motions seem degenerate when the effective mass vanishes.

The physical interpretation of these two additional charges in the plane remains to be studied. However, as we have mentioned already, the symplectic term associated with the second charge $q_2$ is present in other theories in the plane, for instance for the Poincaré group \cite{Feher86} and for the Galilei group \cite{DuvalH01}. In these other examples, it seems to be linked to non commutative coordinates in the plane. The first charge $q_1$ is, to the author's knowledge, not present in other planar theories, and seems specific to the Carrollean case. From the presymplectic forms, \eg \eqref{sigma_2d_em}, or the effective mass \eqref{eff_mass_em}, it seems to be some kind of intrinsic magnetic field (times an electric charge). 

Such charges could have interesting physical consequences. Indeed, as we have recalled, Carroll structures are common in General Relativity. For instance, a photon ``trapped'' on a Kerr-Newman horizon (\ie emitted radially outward right on the horizon) would be subject to the drifting motion shown in \eqref{eom_2d_em_photon}. This motion would be orthogonal to the gradient of the magnetic field on the horizon (in the coordinates \eqref{BHH}), which would be a ``rotating'' effect on top of the frame-dragging effect.

In section \ref{ss:grav} we investigated the gravitational coupling by considering a presymplectic 2-form defined on a curved Cartan geometry, instead of on a flat geometry, which is the case for the free case. We showed that motions remain trivial if spacetime is curved, in the sense that the Riemann tensor does not vanish. However, we found that, in the Cartan geometry, there is room to introduce two additional curvature terms, which are curvature terms associated to the extension part of the Carroll algebra, and hence have no relation with the Riemann tensor. These 2-forms, just like the electromagnetic tensor, may then be decomposed into a magnetic and an electric part. The magnetic components do not change the dynamics by themselves, they are merely equivalent to a shift in the definition of the extensions parameters $q_1$ and $q_2$. The electric tensor associated to these terms however imply non trivial motions. Formally, the expression of the equation of motions in this case is the same as that of the coupling to electromagnetism, in that the Carrollean velocity is orthogonal to the source vector. The physical interpretation of these additional curvature terms remains to be understood, however. 

Last, but not least, we have derived the free quantum equation \eqref{carroll_quant_3} that Carroll wave functions should obey in $d+1$ dimensions, $d \geq 3$. This equation is rather trivial as it does not involve spatial derivatives. This is coherent with classical motions where particles do not move. It is also coherent with \cite{Henneaux79} where it was argued that any Carroll-invariant field theory (based on an action principle) satisfying Carroll causality may only contain derivatives in time.

It is interesting to note that this equation is similar to an equation that appears when one writes the Schr\"odinger equation in a covariant way on a Bargmann structure \cite{DuvalBKP85}. It is then decomposed into two equations, one of which being \eqref{carroll_quant_3}. This is not really a surprise, however, as the slices $t = \const$ of Bargmann structures are Carroll structures, as already mentioned in the introduction (see figure \ref{f:barg}). 

In a recent paper \cite{PetroniloUS21} it was claimed that both an ``extended Carroll Group'' and a ``Klein-Gordon-like equation with Carrollian symmetry'' were obtained. However, it is clear when reading the paper that the obtained ``extension'' of the Carroll group is nothing but the well-known Bargmann group, \ie the central extension of the Galilei group. One can convinced oneself by a direct comparison of the algebra given in \cite[equation $(6)$]{PetroniloUS21} with, \textit{e.g.}, \cite[\S III.B]{LevyLeblond71} upon the redefinitions $C \rightarrow K$, $P_4 \rightarrow -M$, $P_5 \rightarrow -H$. Then, the authors of \cite{PetroniloUS21} construct their equations on a $3+2$ Lorentzian manifold with Bargmann symmetry, and claim they have obtained  a ``non relativistic Klein-Gordon-like equation with Carrollian symmetry''. This is clearly a Bargmann structure, and their equation \cite[equation $(10)$]{PetroniloUS21} is merely the Schr\"odinger equation written in a covariant form, as was already shown in \cite{DuvalBKP85}\footnote{Likewise in the spinor representation, their equations \cite[equation $(18)$]{PetroniloUS21} are merely the well-known Lévy-Leblond equations for non-relativistic (Galilean) spinors written on a Bargmann structure, see \cite{LevyLeblond67,Duval85,DuvalHP96}.}.

Quantization of the planar model remains to be done, but the ``twisted'' aspect of the symplectic form \eqref{2form_ext} complicates the process.

\subsection*{Acknowledgements}

Thanks to Serge Lazzarini for comments and for a careful reading of the manuscript. We are indebted to Peter Horv\'athy and Francisco Herranz for enlighting discussions, and Kevin Morand for useful comments.

The project leading to this publication has received funding from the Excellence Initiative of Aix-Marseille University - A*Midex, a French ``Investissements d’Avenir programme'' AMX-19-IET-008 and AMX-19-IET-009.


\begin{thebibliography}{10}

\bibitem{BacryLL68}
H.~Bacry, J.-M. L{\'e}vy-Leblond, ``{Possible kinematics}'',
  \href{https://dx.doi.org/10.1063/1.1664490 }{ J. Math. Phys. {\bf 9},
  p.~1605, (1968)}.

\bibitem{LevyLeblond76}
J.-M. L{\'e}vy-Leblond, ``{One more derivation of the Lorentz
  transformation}'', \href{https://dx.doi.org/10.1119/1.10490 }{ {Am. J. Phys.}
  {\bf 44}, p.~271, (1976)}.

\bibitem{LevyLeblond65}
J.-M. L\'evy-Leblond, ``Une nouvelle limite non-relativiste du groupe de
  Poincar\'e'', Annales de l'I.H.P. Physique th\'eorique {\bf 3}, p.~1 (1965),
  \url{http://www.numdam.org/item/AIHPA_1965__3_1_1_0}.

\bibitem{LevyLeblond71}
J.-M. L{\'e}vy-Leblond, ``Galilei Group and Galilean Invariance'',
  \href{https://dx.doi.org/10.1016/B978-0-12-455152-7.50011-2 }{ {Loebl Ed.,
  {\bf II}, Acad. Press., New York}, p.~221, (1971)}.

\bibitem{Henneaux79}
M.~Henneaux, ``{Geometry of Zero Signature Space-times}'', Bull. Soc. Math.
  Belg. {\bf 31}, p.~47 (1979).

\bibitem{DuvalGH91}
C.~Duval, G.~W. Gibbons, P.~A. Horv{\'a}thy, ``{Celestial mechanics, conformal
  structures and gravitational waves}'',
  \href{https://dx.doi.org/10.1103/PhysRevD.43.3907 }{ Phys. Rev. D {\bf 43},
  p.~3907, (1991)}, arXiv:
  \href{https://arxiv.org/abs/hep-th/0512188}{hep-th/0512188}.

\bibitem{Dautcourt98}
G.~Dautcourt, ``{On the ultrarelativistic limit of general relativity}'', Acta
  Phys. Polon. B {\bf 29}, p.~1047 (1998), arXiv:
  \href{https://arxiv.org/abs/gr-qc/9801093}{gr-qc/9801093}.

\bibitem{DuvalGH14}
C.~Duval, G.~W. Gibbons, P.~A. Horv{\'a}thy, ``{Conformal Carroll groups and
  BMS symmetry}'', \href{https://dx.doi.org/10.1088/0264-9381/31/9/092001 }{
  Classical Quantum Gravity {\bf 31}, p.~092001, (2014)}, arXiv:
  \href{https://arxiv.org/abs/1402.5894}{1402.5894}.

\bibitem{DuvalGHZ14}
C.~Duval, G.~W. Gibbons, P.~A. Horv{\'a}thy, P.~M. Zhang, ``{Carroll versus
  Newton and Galilei: two dual non-Einsteinian concepts of time}'',
  \href{https://dx.doi.org/10.1088/0264-9381/31/8/085016 }{ Class. Quant. Grav.
  {\bf 31}, p.~085016, (2014)}, arXiv:
  \href{https://arxiv.org/abs/1402.0657}{1402.0657}.

\bibitem{Cartan23}
E.~Cartan, ``{Sur les vari\'et\'es \`a connexion affine et la th\'eorie de la
  relativit\'e g\'en\'eralis\'ee. (première partie)}'', Annales Sci. Ecole
  Norm. Sup. {\bf 40}, p.~325 (1923),
  \url{http://www.numdam.org/item/ASENS_1923_3_40__325_0}.

\bibitem{Trautman63}
A.~Trautman, ``{Sur la th\'eorie newtonienne de la gravitation}'', C.R. Acad.
  Sci. Paris {\bf 257}, p.~617 (1963),
  \url{https://gallica.bnf.fr/ark:/12148/bpt6k4007z/f639.image}.

\bibitem{Havas64}
P.~Havas, ``Four-Dimensional Formulations of Newtonian Mechanics and Their
  Relation to the Special and the General Theory of Relativity'',
  \href{https://dx.doi.org/10.1103/RevModPhys.36.938 }{ Rev. Mod. Phys. {\bf
  36}, p.~938, (1964)}.

\bibitem{Trautman67}
A.~Trautman, ``Comparison of Newtonian and relativistic theories of
  space-time'', Perspectives in Geometry and Relativity. Hoffmann, Banesh
  (ed.). Bloomington, Ind., Indiana University Press, 1966., p.~413 (1967),
  \url{http://trautman.fuw.edu.pl/publications/Papers-in-pdf/22.pdf}.

\bibitem{Kunzle72}
H.~P. K{\"u}nzle, ``{Galilei and lorentz structures on space-time - comparison
  of the corresponding geometry and physics}'', Ann. Inst. H. Poincare Phys.
  Theor. {\bf 17}, p.~337 (1972),
  \url{http://www.numdam.org/item/AIHPA_1972__17_4_337_0}.

\bibitem{Morand18}
K.~Morand, ``{Embedding Galilean and Carrollian geometries I. Gravitational
  waves}'', \href{https://dx.doi.org/10.1063/1.5130907 }{ J. Math. Phys. {\bf
  61}, p.~082502, (2020)}, arXiv:
  \href{https://arxiv.org/abs/1811.12681}{1811.12681}.

\bibitem{CiambelliLMP19}
L.~Ciambelli, R.~G. Leigh, C.~Marteau, P.~M. Petropoulos, ``{Carroll
  Structures, Null Geometry and Conformal Isometries}'',
  \href{https://dx.doi.org/10.1103/PhysRevD.100.046010 }{ Phys. Rev. D {\bf
  100}, p.~046010, (2019)}, arXiv:
  \href{https://arxiv.org/abs/1905.02221}{1905.02221}.

\bibitem{DonnayM19}
L.~Donnay, C.~Marteau, ``{Carrollian Physics at the Black Hole Horizon}'',
  \href{https://dx.doi.org/10.1088/1361-6382/ab2fd5 }{ Class. Quant. Grav. {\bf
  36}, p.~165002, (2019)}, arXiv:
  \href{https://arxiv.org/abs/1903.09654}{1903.09654}.

\bibitem{Ashtekar14}
A.~Ashtekar, ``{Geometry and Physics of Null Infinity}'', arXiv:
  \href{https://arxiv.org/abs/1409.1800}{1409.1800}.

\bibitem{BondiBM62}
H.~Bondi, M.~G.~J. van~der Burg, A.~W.~K. Metzner, ``{Gravitational waves in
  general relativity. 7. Waves from axisymmetric isolated systems}'',
  \href{https://dx.doi.org/10.1098/rspa.1962.0161 }{ Proc. Roy. Soc. Lond. A
  {\bf 269}, p.~21, (1962)}.

\bibitem{Sachs62}
R.~Sachs, ``{Asymptotic symmetries in gravitational theory}'',
  \href{https://dx.doi.org/10.1103/PhysRev.128.2851 }{ Phys. Rev. {\bf 128},
  p.~2851, (1962)}.

\bibitem{Bargmann54}
V.~Bargmann, ``{On Unitary ray representations of continuous groups}'',
  \href{https://dx.doi.org/10.2307/1969831 }{ Annals of Mathematics {\bf 59},
  p.~1, (1954)}.

\bibitem{DuvalBKP85}
C.~Duval, G.~Burdet, H.~P. K{\"u}nzle, M.~Perrin, ``{Bargmann Structures and
  Newton-cartan Theory}'', \href{https://dx.doi.org/10.1103/PhysRevD.31.1841 }{
  Phys. Rev. D {\bf 31}, p.~1841, (1985)}.

\bibitem{Eisenhart28}
L.~P. Eisenhart, ``Dynamical Trajectories and Geodesics'',
  \href{https://dx.doi.org/10.2307/1968307 }{ Annals of Mathematics {\bf 30},
  p.~591, (1928)}.

\bibitem{DuvalGHZ17}
C.~Duval, G.~W. Gibbons, P.~A. Horv{\'a}thy, P.~M. Zhang, ``{Carroll symmetry
  of plane gravitational waves}'',
  \href{https://dx.doi.org/10.1088/1361-6382/aa7f62 }{ Class. Quant. Grav. {\bf
  34}, p.~175003, (2017)}, arXiv:
  \href{https://arxiv.org/abs/1702.08284}{1702.08284}.

\bibitem{BergshoeffGL14}
E.~Bergshoeff, J.~Gomis, G.~Longhi, ``{Dynamics of Carroll Particles}'',
  \href{https://dx.doi.org/10.1088/0264-9381/31/20/205009 }{ Class. Quant.
  Grav. {\bf 31}, p.~205009, (2014)}, arXiv:
  \href{https://arxiv.org/abs/1405.2264}{1405.2264}.

\bibitem{Trzesniewski18}
T.~Trze\'sniewski, ``{Effective Chern\textendash{}Simons actions of particles
  coupled to 3D gravity}'',
  \href{https://dx.doi.org/10.1016/j.nuclphysb.2018.01.023 }{ Nucl. Phys. B
  {\bf 928}, p.~448, (2018)}, arXiv:
  \href{https://arxiv.org/abs/1706.01375}{1706.01375}.

\bibitem{BagchiBKM16}
A.~Bagchi, R.~Basu, A.~Kakkar, A.~Mehra, ``{Flat Holography: Aspects of the
  dual field theory}'', \href{https://dx.doi.org/10.1007/JHEP12(2016)147 }{
  JHEP {\bf 12}, p.~147, (2016)}, arXiv:
  \href{https://arxiv.org/abs/1609.06203}{1609.06203}.

\bibitem{CardonaGP16}
B.~Cardona, J.~Gomis, J.~M. Pons, ``{Dynamics of Carroll Strings}'',
  \href{https://dx.doi.org/10.1007/JHEP07(2016)050 }{ JHEP {\bf 07}, p.~050,
  (2016)}, arXiv: \href{https://arxiv.org/abs/1605.05483}{1605.05483}.

\bibitem{NgendakumanaNT14}
A.~Ngendakumana, J.~Nzotungicimpaye, L.~Todjihounde, ``{Group theoretical
  construction of planar Noncommutative Phase Spaces}'',
  \href{https://dx.doi.org/10.1063/1.4862843 }{ J. Math. Phys. {\bf 55},
  p.~013508, (2014)}, arXiv: \href{https://arxiv.org/abs/1308.3065}{1308.3065}.

\bibitem{Ngendakumana14}
A.~Ngendakumana, {\em {Group Theoretical Construction of Planar Noncommutative
  Systems}}.
\newblock PhD thesis, Université d’Abomey Calavi (2014).
\newblock arXiv: \href{https://arxiv.org/abs/1401.5213}{1401.5213}.

\bibitem{AzcarragaHPS98}
J.~A. de~Azcarraga, F.~J. Herranz, J.~C. Perez~Bueno, M.~Santander, ``{Central
  extensions of the quasiorthogonal Lie algebras}'',
  \href{https://dx.doi.org/10.1088/0305-4470/31/5/008 }{ J. Phys. A {\bf 31},
  p.~1373, (1998)}, arXiv:
  \href{https://arxiv.org/abs/q-alg/9612021}{q-alg/9612021}.

\bibitem{BallesterosGD92}
A.~Ballesteros, M.~Gadella, M.~A. Del~Olmo, ``{Moyal quantization of 2+1
  dimensional Galilean systems}'', \href{https://dx.doi.org/10.1063/1.529939 }{
  J. Math. Phys. {\bf 33}, p.~3379, (1992)}.

\bibitem{InonuW53}
E.~Inonu, E.~P. Wigner, ``On the Contraction of Groups and Their
  Representations'', \href{https://dx.doi.org/10.1073/pnas.39.6.510 }{
  Proceedings of the National Academy of Sciences {\bf 39}, p.~510, (1953)}.

\bibitem{Marsot16}
L.~Marsot, ``Caract{\'e}risation g{\'e}om{\'e}trique des structures de Bargmann
  et de Carroll et des groupes de Schr\"odinger et de Bondi-Metzner-Sachs'',
  Master's thesis, Aix-Marseille University, 2016.
\newblock In french.

\bibitem{Leinaas77}
J.~M. Leinaas, J.~Myrheim, ``{On the theory of identical particles}'',
  \href{https://dx.doi.org/10.1007/BF02727953 }{ Nuovo Cim. B {\bf 37}, p.~1,
  (1977)}.

\bibitem{Wilczek82}
F.~Wilczek, ``{Quantum Mechanics of Fractional Spin Particles}'',
  \href{https://dx.doi.org/10.1103/PhysRevLett.49.957 }{ Phys. Rev. Lett. {\bf
  49}, p.~957, (1982)}.

\bibitem{Kirillov62}
A.~A. Kirillov, ``{Unitary representations of nilpotent Lie groups}'',
  \href{https://dx.doi.org/10.1070/RM1962v017n04ABEH004118 }{ Russ. Math. Surv.
  {\bf 17}, p.~57, (1962)}.

\bibitem{Souriau70}
J.-M. Souriau, {\em Structure des syst\`emes dynamiques}.
\newblock Dunod, Paris, 1970, doi:
  \href{https://doi.org/10.1007/978-1-4612-0281-3}{10.1007/978-1-4612-0281-3}.
\newblock Translation to English: \emph{Structure of Dynamical Systems. A
  Symplectic View of Physics.} (Birkh\"auser, Basel, 1997).

\bibitem{DuvalH01}
C.~Duval, P.~A. Horv{\'a}thy, ``{Exotic Galilean symmetry in the noncommutative
  plane, and the Hall effect}'',
  \href{https://dx.doi.org/10.1088/0305-4470/34/47/314 }{ J. Phys. A {\bf 34},
  p.~10097, (2001)}, arXiv:
  \href{https://arxiv.org/abs/hep-th/0106089}{hep-th/0106089}.

\bibitem{Feher86}
L.~Feh{\'e}r, ``On the coadjoint orbits of the planar Poincar{\'e} group'',

\bibitem{HenneauxS21}
M.~Henneaux, P.~Salgado-Rebolledo, ``{Carroll contractions of Lorentz-invariant
  theories}'', arXiv: \href{https://arxiv.org/abs/2109.06708}{2109.06708}.

\bibitem{BoerHOSV21}
J.~de~Boer, J.~Hartong, N.~A. Obers, W.~Sybesma, S.~Vandoren, ``{Carroll
  symmetry, dark energy and inflation}'', arXiv:
  \href{https://arxiv.org/abs/2110.02319}{2110.02319}.

\bibitem{DuvalGH14b}
C.~Duval, G.~W. Gibbons, P.~A. Horvathy, ``{Conformal Carroll groups}'',
  \href{https://dx.doi.org/10.1088/1751-8113/47/33/335204 }{ J. Phys. A {\bf
  47}, p.~335204, (2014)}, arXiv:
  \href{https://arxiv.org/abs/1403.4213}{1403.4213}.

\bibitem{Kun72}
H.-P. K\"unzle, ``Canonical Dynamics of Spinning Particles in Gravitational and
  Electromagnetic Fields'',
  \href{https://dx.doi.org/https://doi.org/10.1063/1.1666045 }{ J. Math. Phys.
  {\bf 13}, p.~739, (1972)}.

\bibitem{TheseChristian}
C.~Duval, {\em Quelques procédures géométriques en dynamique des particles}.
\newblock PhD thesis, Université Aix-Marseille II (1982).

\bibitem{KobayashiN63}
S.~Kobayashi, K.~Nomizu, {\em Foundations of Differential Geometry}.
\newblock Wiley Classics Library, 1963.

\bibitem{Kostant69}
B.~Kostant, ``{On certain unitary representations which arise from a
  quantization theory}'', \href{https://dx.doi.org/10.1007/3-540-05310-7\_28 }{
  Conf. Proc. C {\bf 690722}, p.~237, (1969)}.

\bibitem{Niederer72}
U.~Niederer, ``{The maximal kinematical invariance group of the free
  Schrodinger equation.}'', \href{https://dx.doi.org/10.5169/seals-114417 }{
  Helv. Phys. Acta {\bf 45}, p.~802, (1972)}.

\bibitem{PetroniloUS21}
G.~X.~A. Petronilo, S.~C. Ulhoa, A.~E. Santana, ``{Representations of Extended
  Carroll Group}'', \href{https://dx.doi.org/10.1007/s00006-021-01146-3 }{ Adv.
  Appl. Clifford Algebras {\bf 31}, p.~41, (2021)}, arXiv:
  \href{https://arxiv.org/abs/2104.12535}{2104.12535}.

\bibitem{LevyLeblond67}
J.-M. L{\'e}vy-Leblond, ``{Nonrelativistic particles and wave equations}'',
  \href{https://dx.doi.org/10.1007/BF01646020 }{ Commun. Math. Phys. {\bf 6},
  p.~286, (1967)}.

\bibitem{Duval85}
C.~Duval, ``{The Dirac and Levy-Leblond Equations and Geometric
  Quantization}'', \href{https://dx.doi.org/10.1007/BFb0077322 }{ Lect. Notes
  Math. {\bf 1251}, p.~205, (1987)}.

\bibitem{DuvalHP96}
C.~Duval, P.~A. Horv{\'a}thy, L.~Palla, ``{Spinors in nonrelativistic
  Chern-Simons electrodynamics}'',
  \href{https://dx.doi.org/10.1006/aphy.1996.0071 }{ Annals Phys. {\bf 249},
  p.~265, (1996)}, arXiv:
  \href{https://arxiv.org/abs/hep-th/9510114}{hep-th/9510114}.

\end{thebibliography}

\end{document}